\documentclass{aa}
\usepackage{graphicx}
\usepackage{txfonts}

\begin{document}

\title{Chemical differentiation in regions of high-mass star formation}
\subtitle{I. CS, dust and N$_2$H$^+$ in southern sources}

\author{L. Pirogov \inst{1}
\and I. Zinchenko \inst{1,2,3}
\and P. Caselli \inst{4,5}
\and L. E. B. Johansson \inst{6}}

\offprints{L.~Pirogov \email{pirogov@appl.sci-nnov.ru}}

\institute{
Institute of Applied Physics of the Russian Academy of Sciences, Ulyanova 46,
603950 Nizhny Novgorod, Russia
\and Nizhny Novgorod University , Gagarin av. 23,
603950 Nizhny Novgorod, Russia
\and Helsinki University Observatory, T\"ahtitorninm\"aki, P.O. Box 14, FIN-00014
University of Helsinki, Finland
\and INAF -- Osservatorio Astrofisico di Arcetri, Largo E. Fermi 5, I-50125 Firenze,
Italy
\and Harvard-Smithsonian Center for Astrophysics, 60 Garden St., Cambridge,
MA 02138, USA
\and Onsala Space Observatory, S-43992, Onsala, Sweden}

\date{Received / Accepted}

\abstract
{}
{Our goals are to compare the CS, N$_2$H$^+$ and dust distributions
in a representative sample of high-mass star
forming dense cores and to determine the physical and chemical
properties of these cores.}
{We compare the results of CS(5--4) and 1.2~mm continuum mapping
of twelve dense cores from the southern hemisphere presented in
this work, in combination with our previous N$_2$H$^+$(1--0) and CS(2--1) data.
We use numerical modeling of molecular excitation to estimate
physical parameters of the cores.}
{Most of the maps have several emission peaks (clumps).
Mean sizes of 17 clumps, having counterparts in continuum and CS,
are 0.30(0.06)~pc (continuum) and 0.51(0.07)~pc (CS).
For the clumps with IRAS sources we derived
dust temperatures: 24--35~K, masses: 90--6900~$M_{\odot}$,
molecular hydrogen column densities: $(0.7-12.0)\times 10^{23}$~cm$^{-2}$
and luminosities: ($0.6-46.0)\times 10^4~L_{\odot}$.
LVG densities towards CS peaks within the 50$''$ beam (0.56~pc at 2.3~kpc,
the average distance of our sample source) vary from source to source
in the range: (3--40)$\times$\,$10^5$~cm$^{-3}$.
Masses calculated from LVG densities are higher than CS virial masses
and masses derived from continuum data, implying small-scale clumpiness
of the cores.
The molecular abundances towards IRAS sources in eight objects
are $X$(CS)=$(0.3-2.7)\times10^{-9}$
and $X$(N$_2$H$^+$)=$(0.3-4.4)\times10^{-10}$.
The CS and continuum maps have been compared with each other and
with the N$_2$H$^+$(1--0) maps.
For most of the objects, the CS and continuum peaks are close to
the IRAS point source positions.
The CS(5--4) intensities correlate with continuum fluxes per beam in all
cases, but only in five cases with the N$_2$H$^+$(1--0) intensities.
The study of spatial variations of molecular integrated intensity ratios
to continuum fluxes per beam reveals that
$I$(N$_2$H$^+$)/$F_{1.2}$ ratios drop towards the CS peaks for most
of the sources, which can be due to a N$_2$H$^+$ abundance
decrease.
For CS(5--4), the $I$(CS)/$F_{1.2}$ ratios show no clear trends with distance
from the CS peaks, while for CS(2--1) such ratios drop towards these peaks.
Possible explanations of these results are considered.
The analysis of normalized velocity differences between CS and N$_2$H$^+$ lines
has not revealed indications of systematic motions towards CS peaks.
}
{}


\authorrunning{L. Pirogov et al.}

\titlerunning{Chemical differentiation in regions of high-mass star formation I.}

\maketitle

\section{Introduction}

The first step in most studies of dense cores of molecular clouds
is their identification, either by a visual inspection
or by rather sophisticated numerical analysis of molecular line maps.
Yet, the maps of ``traditional" tracers of dense gas
(CS, HCN, HCO$^+$, NH$_3$, N$_2$H$^+$ etc.)
appear to be different from each other in many cases.
Most distinct differences are often observed between CS and N$_2$H$^+$ maps.
In low-mass starless cores (e.g. Tafalla et al. 2002) the CS emission
vanishes towards the core center while the N$_2$H$^+$ emission is still high
in the interior regions.
Some more turbulent starless clumps show strong CS emission in comparison
with N$_2$H$^+$ (e.g. Williams \& Myers 1999, Olmi et al. 2005).
The optically thick CS lines in the latter case are usually asymmetric
and blueshifted, implying that they are probably
tracing infall motions.
Both these cases can be explained by time dependent chemical models
including depletion molecular species onto grains
(Bergin \& Langer 1997, Bergin et al. 1997, Li et al. 2002, Shematovich et al.
2003, Aikawa et al. 2003), suggesting different evolutionary
stages of clumps within parent molecular cloud.
In the cores with embedded stellar objects, especially those of high mass,
the morphologies of molecular maps become more complicated. This is due
to the overlapping of the emission coming from the quiescent gas with that
from outflows and hot cores, where dust grain mantles evaporate because
of their proximity to YSOs.
The N$_2$H$^+$ intensities towards these objects may drop,
while CS peaks towards YSO positions
(e.g. Ungerechts et al. 1997, Bottinelli \& Williams 2004).
Thus, chemical models of starless clumps cannot be used to explain
the observed differences between CS and N$_2$H$^+$ maps in high-mass star
forming regions.

In the last decade, a large sample of high-mass star forming cores
associated with water masers
have been mapped in the CS(2--1) line (Zinchenko et al. 1994, 1995, 1998).
Many of these cores have subsequently been mapped in different
molecular lines, including N$_2$H$^+$(1--0) (Pirogov et al. 2003,
hereafter Paper~I). Thus, the maps can be compared and important
quantities can be determined for these regions, such as density
distribution and chemical composition.
Yet, in contrast to the N$_2$H$^+$(1--0) lines, which are optically thin
in most cases (Paper~I), the CS(2--1) lines are likely optically thick and
possible differences between maps in these lines
could be connected both with optical depth effects and chemical differences.

In this paper we present the results of 1.2~mm dust continuum and CS(5--4)
observations in several southern sources.
Optically thin dust emission is known to trace closely
the gas component in high density regions
and it is insensitive to molecular abundance variations and radiative
transfer effects.
This can be used to separate optical depth and chemical
effects in molecular line maps.

The goal of the paper is to analyze the dust continuum and CS(5--4) data
in comparison with our previous CS(2--1) and N$_2$H$^+$(1--0) results
in order to get reliable information on the density and chemical structure
of the cores.
The paper also contains estimates of physical parameters
including sizes, masses as well as density distributions derived from
CS(5--4) and CS(2--1) data and molecular abundances for several sources.

\section{Observations}

\subsection{Source list}

The sources were originally selected from
the sample of high-mass star forming regions associated with water masers
located in the southern hemisphere (Zinchenko et al. 1995, Juvela 1996)
and from the dense core database of Jijina et al. (1999), according to
the following criteria: presence of embedded clusters of stars detected
in infrared, and distances to the objects not greater than 5~kpc.
If no data on clusters were available,
we selected sources with high IR-luminosities
($L>10^4$ L$_{\odot}$), which can be an indirect indication
of cluster (Jijina et al. 1999).
In total, 14 sources have been selected for CS(5--4) observations,
twelve of them have been observed in continuum at 1.2 mm wavelength.
Most of the sources have associated IRAS point sources.
The source list with coordinates and distances is given
in Table~\ref{table:list}.

\begin{table}[htb]
\centering
\caption[]{Source list}
\begin{tabular}{lrrc}
\noalign{\hrule}\noalign{\smallskip}
Source
          & RA (2000)                      & Dec (2000)                 & $D$   \\
          & ${\rm (^h)\  (^m)\  (^s)\ }$ &($\degr$ $\arcmin$ $\arcsec$) & (kpc) \\
\noalign{\smallskip}\hline\noalign{\smallskip}
G~264.28$+$1.48  &08 56 27.8   &$-$43 05 05  & 1.4$^a$ \\
G~265.14$+$1.45  &08 59 24.7   &$-$43 45 22  & 1.7$^a$ \\
G~267.94$-$1.06  &08 59 03.6   &$-$47 30 47  & 0.7$^a$ \\
G~268.42$-$0.85  &09 01 54.3   &$-$47 43 59  & 1.3$^a$ \\
G~269.11$-$1.12  &09 03 32.8   &$-$48 28 39  & 2.6$^a$ \\
G~270.26$+$0.83  &09 16 43.3   &$-$47 56 36  & 2.6$^a$ \\
G~285.26$-$0.05  &10 31 30.0   &$-$58 02 07  & 4.7$^a$ \\
G 291.27$-$0.71  &11 11 49.9   &$-$61 18 14  & 2.7$^b$ \\
G~294.97$-$1.73  &11 39 12.6   &$-$63 28 47  & 1.2$^a$ \\
G~305.36$+$0.15  &13 12 33.9   &$-$62 37 38  & 4.2$^a$ \\
G~316.77$-$0.02  &14 44 58.9   &$-$59 48 29  & 3.1$^c$ \\
G~345.01$+$1.80  &16 56 45.3   &$-$40 14 03  & 2.1$^c$ \\
G~345.41$-$0.94  &17 09 33.7   &$-$41 35 52  & 2.8$^c$ \\
G~351.41$+$0.64  &17 20 53.4   &$-$35 47 00  & 1.7$^d$ \\
\noalign{\smallskip}\hline\noalign{\smallskip}
\end{tabular}

$^a$ Zinchenko et al. (\cite{zin3}),
$^b$ Brand \& Blitz (\cite{brand}),
$^c$ Juvela (\cite{juvela96}),

$^d$ Neckel (\cite{neckel})

\label{table:list}
\end{table}

\normalsize

\subsection{The CS(5--4) observations}

Observations of the CS(5--4) line at 244.9~GHz towards
14 southern sources have been carried out with the 15-m SEST antenna
on February, 2001.
The N$_2$H$^+$(1--0) line at 93.2~GHz was observed simultaneously
and the results on this line are reported in Paper~I.
The telescope half power beam width at the CS(5--4) frequency
is about 22$''$, and the main beam efficiency is 0.5.
The system temperature was $\sim 400-1500$~K,
depending on source elevation and weather conditions.
Spectral analysis was done using an acousto-optical spectrum analyzer
(2000 channels) splitted into two halves to measure the N$_2$H$^+$(1--0)
and the CS(5--4) lines simultaneously.
The frequency resolution was 42.6~kHz which corresponds
to a velocity resolution of 0.052~km~s$^{-1}$ at the CS(5--4) frequency.

Pointing was regularly checked by SiO maser observations and
typically it was better than 5$\arcsec$.
Mapping has been done with 20$\arcsec$ grid spacing.
The data processing included baseline subtraction (low-order polynomials)
and Gaussian fitting.

\subsection{Observations of dust continuum emission}

In June 2003, twelve sample sources have been mapped
in dust continuum emission at 250~GHz with the SEST antenna.
G267.94 and G305.36 were excluded from the list of continuum observations
because they were not completed in CS(5--4) and N$_2$H$^+$(1--0).

We used the 37 elements SIMBA bolometer array.
The HPBW of a single element is about 24$\arcsec$, the separation between
elements on the sky is 44$\arcsec$.
Mapping has been done by scanning in azimuth, sampling the signal every
8$''$. The scans were spaced by 8$''$ in elevation.
This gave the pixel size 8$''$.
Typical map sizes are $900''\times 1200''$.
The raw data have been converted by the $simbaread$ software and the
processing has been done using the MOPSI package written by R.~Zylka
according to the instructions of the SIMBA Observer's Handbook (2003)
\footnote{http://puppis.ls.eso.org/staff/simba/manual/simba/index.html}.
The final r.m.s. noise level derived from regions on the maps without sources
is 30-60~mJy~beam$^{-1}$ for 9 sources.
For three sources (G345.01, G345.41 and G351.41) the r.m.s. noise
level is $160-180$~mJy~beam$^{-1}$.
The flux measurements in the final coadded map are believed
to be correct within 20\%.
Note that the continuum fluxes also include contribution
from molecular lines lying within bolometer bandwidth (mainly CO(2--1)).
Using the Braine et al. (1995) analysis combined with CO(1--0) intensities
(Zinchenko et al. 1995), we estimate that such a contribution is not
significant for any source in our sample.

\section{Results}

\subsection{Maps and line parameters}

The CS(5--4) integrated intensity maps together with continuum
maps are shown in Fig.~\ref{maps}.
For comparison, the N$_2$H$^+$(1--0) maps (Paper~I) are also plotted.
In general, the continuum and the CS(5--4) maps have similar
structures and their morphologies range from
close to spherical-symmetry (e.g. G270.26) to complex clumpy
structures with several emission peaks (e.g. G316.77).
IRAS sources are indicated by stars and the uncertainty ellipses
corresponding to 95\% confidence level in positional accuracy are also shown.
For most of the objects (except G269.11, G316.77 and, probably,
G265.14 and G345.41) these ellipses overlap with the
CS and dust 90\% intensity contours, indicating coincidence
of CS, continuum peaks and IRAS point source positions.

\begin{figure*}[htb]
\centering \includegraphics[width=15cm]{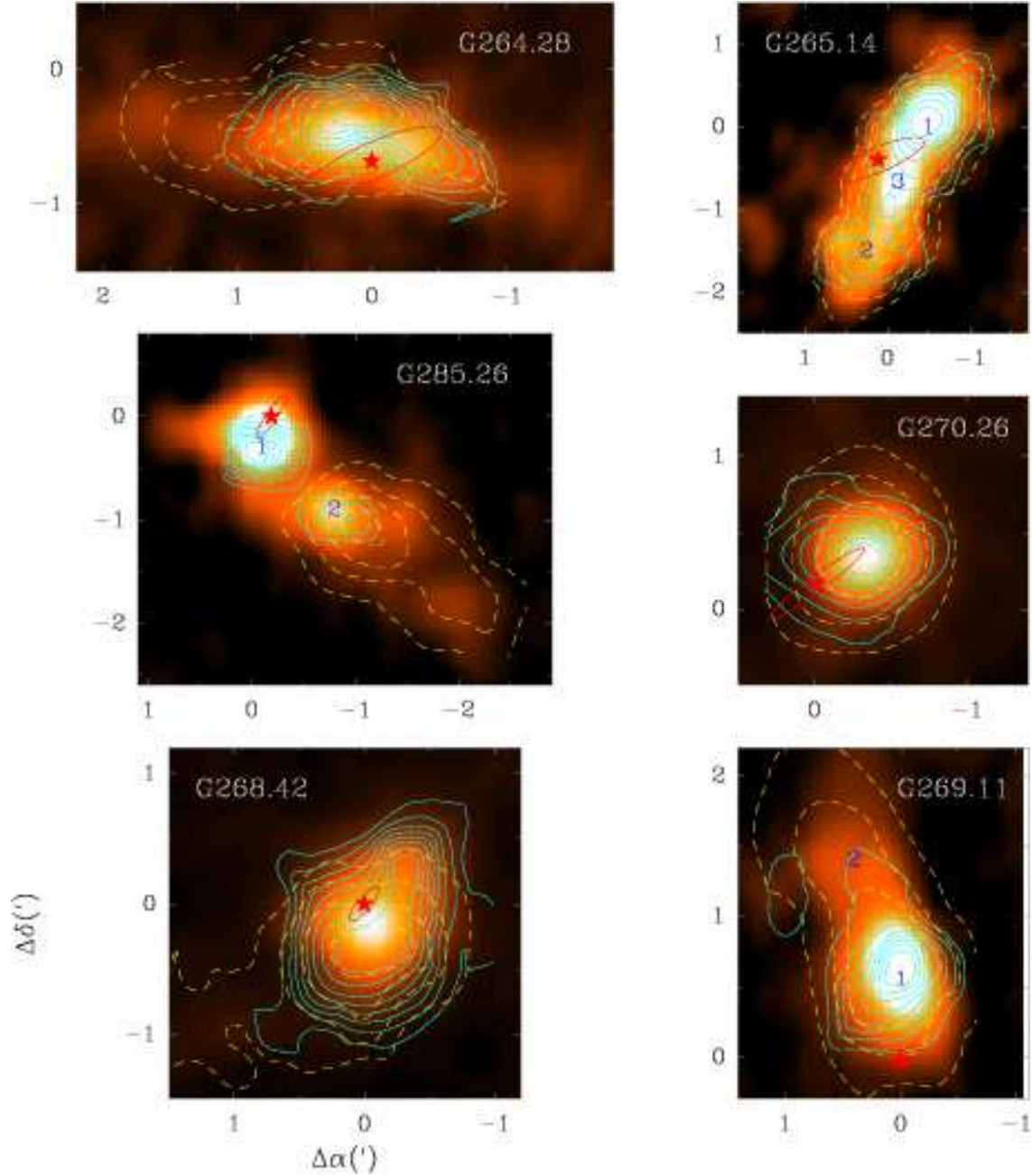}
\caption{Continuum 1.2~mm maps (in color) together with CS(5--4)
(solid blue contours) and N$_2$H$^+$(1--0) (dashed yellow contours,
taken from Paper~I) integrated intensity maps.
Intensity contours range from 10\% to 90\% of peak values
with 10\% step plus 5\% contour in addition (continuum),
from 20\% to 90\% of peak values with 10\% step (CS),
and from 30\% to 90\% of peak values with 20\% step (N$_2$H$^+$).
The CS peak integrated intensities are given in Table~\ref{table:peak}.
The peak continuum fluxes (in Jy beam$^{-1}$) are: 1.46 (G264.28),
2.32 (G265.14), 7.43 (G268.42), 3.44 (G269.11), 3.34 (G270.26),
4.39 (G285.26), 18.83 (G291.27), 1.64 (G294.97), 3.37 (G316.77),
3.93 (G345.01), 3.39 (G345.41), 12.54 (G351.41).
The continuum clumps are marked by numbers as in Table~\ref{clumps2}.
IRAS point sources are marked by red stars.
The uncertainty ellipses corresponding to 95\% confidence level
in IRAS point source position are also shown.
The SEST beam size at the CS(5--4) frequency is shown on the G~264.28 map}
\label{maps}
\end{figure*}

\addtocounter{figure}{-1}

\begin{figure*}[htb]
 \centering \includegraphics[width=15cm]{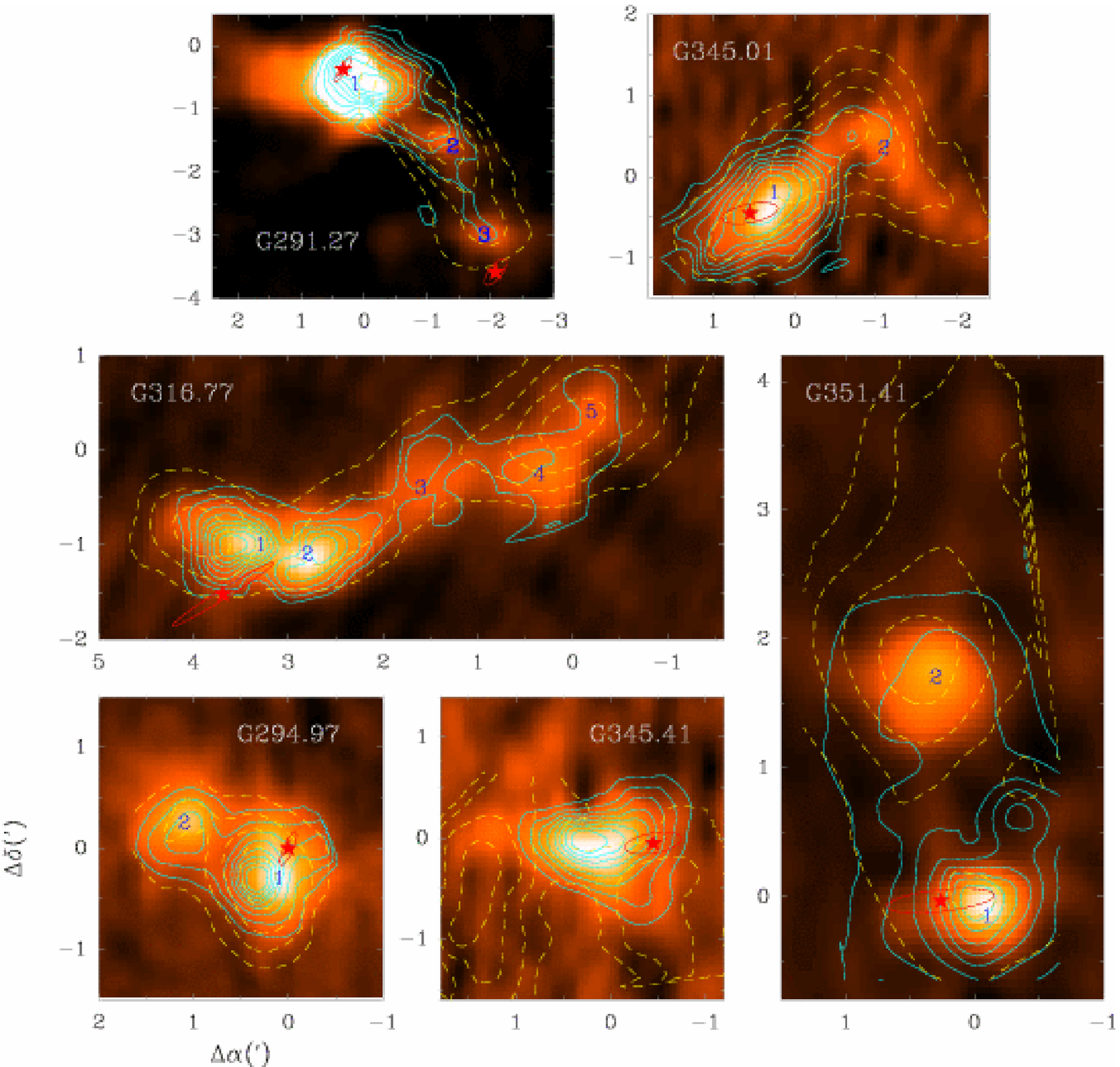}
\caption{continued}
\end{figure*}

Offset coordinates of CS(5--4) integrated intensity peaks,
the values of integrated intensities towards these positions
and parameters of Gaussian fits (main beam temperature,
velocity and line width) are given in Table~\ref{table:peak}
with the corresponding 1~$\sigma$ errors, as defined by the
fits, in brackets.

\begin{table*}[htb]
\centering
\caption[]{The CS(5--4) line parameters}
\small
\begin{tabular}{lrrrrr}
\noalign{\hrule}\noalign{\smallskip}
Source   &($\Delta \alpha \arcsec$, $\Delta \delta \arcsec$)
         &$I$            & $T_{\rm MB}$  & $V_{\rm LSR}$  & $\Delta V$    \\ 
        &&(K km s$^{-1}$)&  (K)          &  (km s$^{-1}$) &  (km s$^{-1}$)\\ 
        &&               &               &                &               \\ 
\noalign{\smallskip}\hline\noalign{\smallskip}

G~264.28$+$1.48 & (0,--20)  & 13.3(0.2) & 5.08(0.06)   &   5.93(0.01) & 2.34(0.03)\\   
G~265.14$+$1.45 & (--20,0)  & 20.9(0.6) & 7.34(0.18)   &   7.80(0.03) & 2.58(0.08)\\   
G~267.94$-$1.06 & (0,20)    & 16.8(1.1) & 4.01(0.21)   &   3.33(0.11) & 4.37(0.27)\\   
G~268.42$-$0.85 & (0,--20)  & 30.8(0.7) &10.26(0.22)   &   3.51(0.03) & 2.90(0.07)\\   
G~269.11$-$1.12 & (0,40)    & 35.8(0.7) & 5.82(0.11)   &   9.74(0.05) & 5.78(0.13)\\   
G~270.26$+$0.83 & (--20,20) & 25.3(0.7) & 6.01(0.17)   &   9.67(0.06) & 4.17(0.14)\\   
G~285.26$-$0.05 & (0,--20)  & 31.8(0.3) & 5.76(0.05)   &   3.67(0.02) & 5.04(0.05)\\   
G 291.27$-$0.71 & (--20,--40)& 43.0(0.6) & 10.44(0.14) &--23.75(0.03) & 4.08(0.07)\\   
G~294.97$-$1.73 & (20,--20) & 14.9(0.5) & 4.71(0.17)   & --8.01(0.05) & 2.78(0.12)\\   
G~305.36$+$0.15 & (20,20)   & 18.4(0.5) & 3.66(0.08)   &--38.56(0.05) & 5.46(0.14)\\   
G~316.77$-$0.02 & (220,--60)& 22.0(0.3) & 6.10(0.08)   &--38.44(0.02) & 3.41(0.05)\\   
G~345.01$+$1.80 & (20,--20) & 44.1(0.5) & 8.03(0.08)   &--13.48(0.03) & 5.32(0.06)\\   
G~345.41$-$0.94 & (20,0)    & 69.2(0.5) &13.87(0.10)   &--21.16(0.02) & 4.82(0.04)\\   
G~351.41$+$0.64 & (0,0)     & 116.8(0.6)&17.19(0.08)   & --7.03(0.01) & 6.42(0.04)\\   
\noalign{\smallskip}\hline\noalign{\smallskip}
\end{tabular}
\label{table:peak}
\end{table*}

\subsection{Clump parameters}
\label{sec:clumps}

The CS(5--4) and continuum maps have been deconvolved into individual
clumps using our 2D Gaussian fitting program and the method described
in Paper~I.
No attempts to separate individual clumps have been made in G264.28 where
local emission peaks of nearly equal intensities are located too close
to each other.
We have not processed the CS(5--4) map in G267.94 as it
has not been completed.
The parameters of individual CS(5--4) clumps, including relative coordinates
of clump centers, aspect ratios (the ratios of the extents of the fitted elliptical
Gaussians) and deconvolved angular ($\Delta\Theta$) and linear ($d$) sizes
estimated as geometric mean of the extents of the elliptical Gaussians
at half maximum intensity level are given in Table~\ref{clumps1}
(columns 2--6).
The clumps which belong to the same object are marked by numbers.

In columns 7 and 8 of Table~\ref{clumps1} the CS(5--4)
mean line widths and virial masses are given.
For homogeneous spherically-symmetric clumps with no external
pressure and no magnetic field, virial masses are given by:

$$
M_{\rm vir}(M_{\odot})=105\,\langle\Delta V\rangle^2 \cdot d \hspace{2mm},
$$

\noindent{where $\langle\Delta V\rangle$ is the CS(5--4) mean line width
(in km~s$^{-1}$), defined as the weighted average of line widths
at different positions within half-maximum intensity region;
$d$ is the CS(5--4) emission region size in pc.}
Virial masses calculated according to the above formula
lie in the range $\sim$\,140\,--\,1630~$M_{\odot}$, with a
mean value of $753~M_{\odot}$. We point out that these values can
be overestimated, given that the CS(5--4) lines can be
broadened due to optical depth.
Shirley et al. (2003) who have studied a large sample of high-mass
star forming regions in the northern hemisphere found that CS(5--4) line widths
are on average 1.3 times higher than those of presumably optically thin
C$^{34}$S(5--4) lines.
In addition, if density in the cores decreases outwards, virial mass
should be multiplied by a factor $\frac{3\,(5-2p)}{5\,(3-p)}$, where
$p$ is the power-law index for density radial profile.
For $p$ close to 2 (Paper I) this factor is close to 0.6.

\begin{table*}
\centering
\caption[]{Physical parameters of the CS(5--4) clumps}
\small
\begin{tabular}{lrrrrrrrr}
\noalign{\hrule}\noalign{\smallskip}
Source   & $\Delta\alpha$  & $\Delta\delta$ & Aspect  &$\Delta \Theta$ & $d$  & $\langle\Delta V\rangle$ &$M_{\rm vir}$   & IRAS \\
         &  ($\arcsec$)    &  ($\arcsec$)   & ratio  &  ($\arcsec$)   &(pc)  & (km s$^{-1}$)            & ($M_{\odot}$)   \\
         &                 &                &        &                &      &                          &                 \\
\noalign{\smallskip}\hline\noalign{\smallskip}

G~264.28$+$1.48     &    4(3) & --26(1)    &3.2(0.8) &  42(4) & 0.28(0.03) & 2.2(0.2) & 142 & + \\
G~265.14$+$1.45 (1) & --29(1) &    0(1)    &1.8(0.1)  & 58(2) & 0.48(0.02) & 2.2(0.2) & 238 & + \\
G~265.14$+$1.45 (2) &   12(2) & --96(1)    &1.6(0.2)  & 57(4) & 0.47(0.03) & 1.7(0.1) & 141 & -- \\
G~268.42$-$0.85     &  --6(1) & --12(2)    &1.6(0.1)  & 64(3) & 0.40(0.02) & 2.7(0.2) & 307 & + \\
G~269.11$-$1.12     &    4(2) &   35(2)    &1.2(0.2)  & 46(4) & 0.58(0.05) & 2.9(0.3) & 506 & -- \\
G~270.26$+$0.83     & --15(2) &   20(1)    &1.5(0.2)  & 38(3) & 0.48(0.04) & 2.6(0.3) & 337 & + \\
G~285.26$-$0.05 (1) & --7(1)  & --20(1)    &1.2(0.2)  & 18(2) & 0.42(0.04) & 4.5(0.3) & 868 & + \\
G~285.26$-$0.05 (2) & --49(2) & --58(1)    &1.9(0.3)  & 34(3) & 0.78(0.06) & 3.4(0.2) & 953 & -- \\
G 291.27$-$0.71     & --8(2)  & --31(1)    &1.4(0.1)  & 70(2) & 0.92(0.03) & 4.0(0.1) &1550 & + \\
G~294.97$-$1.73 (1) &  16(3)  & --13(3)    &1.3(0.3)  & 51(5) & 0.29(0.03) & 2.2(0.2) & 152 & + \\
G~294.97$-$1.73 (2) &  54(7)  &    0(3)    &1.6(0.5)  & 63(9) & 0.37(0.06) & 2.4(0.2) & 219 & -- \\
G~305.36$+$0.15     &  15(2)  &   33(3)    & 1.6(0.5) & 37(5) & 0.76(0.11) & 3.7(0.3) &1090 & -- \\
G~316.77$-$0.02 (1) &  216(1) & --59(1)    & 1.1(0.2) & 32(3) & 0.48(0.04) & 3.9(0.3) & 762 & + \\
G~316.77$-$0.02 (2) &  156(2) & --64(2)    & 2.7(0.6) & 32(4) & 0.49(0.06) & 4.4(0.2) & 973 & + \\
G~316.77$-$0.02 (3) &   93(4) & --17(5)    & 2.1(0.6) & 62(9) & 0.93(0.14) & 4.1(0.4) &1630 & -- \\
G~316.77$-$0.02 (4) &   17(3) & --12(2)    & 1.5(0.3) & 57(6) & 0.85(0.09) & 3.3(0.3) & 985 & -- \\
G~316.77$-$0.02 (5) & --14(3) &   38(5)    & 2.2(1.0) & 39(8) & 0.59(0.13) & 3.7(0.3) & 870 & -- \\
G~345.01$+$1.80     &   22(2) & --23(1)    & 1.9(0.1) & 85(2) & 0.87(0.02) & 4.0(0.1) &1450 & + \\
G~345.41$-$0.94     &    4(2) & --2(1)     & 1.8(0.2) & 59(3) & 0.81(0.04) & 3.6(0.2) &1070 & + \\
G~351.41$+$0.64     &  --3(1) & --5(1)     & 1.1(0.1) & 35(2) & 0.29(0.02) & 5.2(0.1) & 814 & + \\
\noalign{\smallskip}\hline\noalign{\smallskip}
\end{tabular}
\label{clumps1}
\end{table*}

The columns 2--6 of Table~\ref{clumps2} contain parameters
of continuum clumps similar to those of CS clumps.
Total fluxes ($F_{\rm total}$) are given in column 7.
A presence or an absence of an IRAS point source within the
half maximum intensity
level of individual clumps is marked by plus or minus signs in the last column
of Table~\ref{clumps1} and Table~\ref{clumps2}.

\begin{table*}
\centering
\caption[]{Physical parameters of dust clumps}
\small
\begin{tabular}{lrrrrrrr}
\noalign{\hrule}\noalign{\smallskip}
Source   & $\Delta\alpha$  & $\Delta\delta$ & Aspect  &$\Delta \Theta$ & $d$ & $F_{\rm total}$ & IRAS\\
         &  ($\arcsec$)    &  ($\arcsec$)   & ratio  &  ($\arcsec$)   &(pc) & (Jy) \\
\noalign{\smallskip}\hline\noalign{\smallskip}

G~264.28$+$1.48     &  2.2(0.4) & --33.0(0.2)   &4.0(0.2)  & 36(1) & 0.24(0.01)      & 5.0 & + \\
G~265.14$+$1.45 (1) &--28.1(0.3)&    3.2(0.3)   &1.4(0.1)  & 30(1) & 0.25(0.01)      & 5.1 & + \\
G~265.14$+$1.45 (2) & 14.4(0.6) &--88.7(0.6)    &1.3(0.1)  & 50(1) & 0.42(0.01)      & 5.8 & -- \\
G~265.14$+$1.45 (3) &--8.1(0.3) &--38.6(0.5)    &1.5(0.1)  & 36(1) & 0.30(0.01)      & 5.0 & + \\
G~268.42$-$0.85     & --5.9(0.2)& --6.9(0.2)    &1.37(0.03)& 38.2(0.4) & 0.241(0.003)&24.8 & + \\
G~269.11$-$1.12 (1) & 1.1(0.2)  &  34.5(0.2)    &1.6(0.1)  & 27(1)     & 0.34(0.01)  & 7.5 & -- \\
G~269.11$-$1.12 (2) & 24(1)     &  82(1)        &1.4(0.1)  & 44(2)     & 0.56(0.03)  & 2.9 & -- \\
G~270.26$+$0.83     &--21.6(0.2)&  20.9(0.2)    &1.2(0.1)  & 20.8(0.4) & 0.26(0.01)  & 6.6 & + \\
G~285.26$-$0.05 (1) & --6.7(0.1)& --11.3(0.1)   &1.03(0.03)& 19.8(0.3) & 0.45(0.01)  & 7.0 & + \\
G~285.26$-$0.05 (2) &--48.3(0.4)& --56.7(0.3)   &1.5(0.4)  & 12(2)     & 0.27(0.03)  & 1.7 & -- \\
G~291.27$-$0.71 (1) &11.0(0.1)  & --36.2(0.1)   &1.28(0.02)& 41.9(0.3) & 0.548(0.004)&74.5 & + \\
G~291.27$-$0.71 (2) &--74(1)    & --92(1)       & 2.5(0.2) & 45(2)     & 0.59(0.03)  & 6.2 & -- \\
G~291.27$-$0.71 (3) &--121(1)   & --183(1)      & 1.1(0.1) & 50(1)     & 0.65(0.02)  & 5.0 & -- \\
G~294.97$-$1.73 (1) &  6.4(0.4) & --12.5(0.4)   & 1.4(0.1) & 44(1)  & 0.254(0.004)   & 5.9 & + \\
G~294.97$-$1.73 (2) &  65(1)    &  15.7(0.4)    & 1.1(0.1) & 42(1)  & 0.24(0.01)     & 3.9 & -- \\
G~316.77$-$0.02 (1) &  200(1)   & --61.8(0.3)   & 2.7(0.1) & 46(1)  & 0.69(0.01)     & 9.1 & + \\
G~316.77$-$0.02 (2) &  168(1)   & --66.2(0.4)   & 2.5(0.2) & 41(1)  & 0.62(0.02)     & 8.7 & + \\
G~316.77$-$0.02 (3) &  95(1)    & --24(1)       & 1.8(0.1) & 56(1)  & 0.84(0.02)     & 6.2 & -- \\
G~316.77$-$0.02 (4) &  21(1)    & --15(1)       & 1.1(0.1) & 40(1)  & 0.60(0.02)     & 5.7 & -- \\
G~316.77$-$0.02 (5) & --14(1)   &   25(1)       & 1.3(0.1) & 39(2)  & 0.59(0.02)     & 3.6 & -- \\
G~345.01$+$1.80 (1) &   15(1)   & --22.0(0.4)   & 2.4(0.1) & 42(1)  & 0.43(0.01)     &14.3 & + \\
G~345.01$+$1.80 (2) & --69(1)   &   22(1)       & 1.9(0.2) & 48(2)  & 0.49(0.02)     & 6.6 & -- \\
G~345.41$-$0.94     &   10(1)   & --7.7(0.4)    & 2.3(0.1) & 42(1)  & 0.57(0.02)     &12.6 & + \\
G~351.41$+$0.64 (1) & --5.2(0.3)  & --5.5(0.2)  & 2.1(0.2) & 16(1)  & 0.13(0.01)     &19.9 & + \\
G~351.41$+$0.64 (2) &  20.3(0.3)  &  99.8(0.3)  &1.05(0.04)& 32(1)  & 0.26(0.01)     &21.8 & -- \\
\noalign{\smallskip}\hline\noalign{\smallskip}
\end{tabular}
\label{clumps2}
\end{table*}

All the CS(5--4) clumps have corresponding counterparts in the continuum.
Their centers coincide within 10$''$ in most cases.
Larger angular distances (up to 16$''$) between CS and continuum clump centers
have been found in G294.97(2) and G316.77(1), however,
in these cases the sensitivity of the CS data is rather low.
No attempts to reveal individual clumps have been done
in the case of G264.28 (both for continuum and CS maps) and G291.27
(for CS map) where the emission peaks are too close to each other
and have comparable intensities.

In several cases (G265.14, G285.26, G291.27, G345.01 and G351.41)
the continuum maps show more detailes than their CS(5--4) counterparts
probably due to a higher signal-to-noise ratio, better spatial sampling
and absence of optical depth effects.
In G351.41, the continuum map reveals two clumps separated by about 110$''$
in the north-south direction which are associated with the well-known massive
star forming regions NGC~6334~I and NGC~6334~I(N), thought to be at
different evolutionary stages (e.g. McCutcheon et al. 2000).
Although the northern clump in our CS(5--4) map is not so prominent
as in continuum, the map shows similar morphology
and is in general agreement with the CS(7--6) results
from McCutcheon et al. (2000).

Angular and linear sizes of 20 CS clumps lie in the ranges:
$\Delta\Theta=18-85''$ and $d=0.28-0.93$~pc
with $\langle\Delta\Theta\rangle=50''(6'')$ and
$\langle d\rangle=0.52(0.07)$~pc; the numbers in brackets are r.m.s. deviations
from the mean.
The sizes of 25 nearby continuum clumps are: $\Delta\Theta=12-56''$ and
$d=0.13-0.84$~pc with $\langle\Delta\Theta\rangle=33(4)''$
and $\langle d\rangle=0.34(0.06)$~pc.
The continuum clumps in most of the sources are smaller than the CS clumps.
For 17 clumps detected both in the continuum and CS
(excluding G264.28 and G291.27) $\langle d\rangle=0.30(0.06)$~pc
(continuum) and 0.51(0.07)~pc (CS).
The difference in sizes could be connected with low
signal-to-noise ratios and undersampling of the CS maps which may produce
enlarged CS clumps sizes.
Higher signal-to-noise ratios in the continuum observations allowed us to reveal
multiple clumps in some sources (e.g. G269.11, G291.27)
while in CS we see single large clump.
Yet, it is also possible that the differences in sizes could be due
to the CS opacity or abundance gradients within clumps.
The CS clump sizes are close to those of N$_2$H$^+$ ones (Paper~I).
Aspect ratios for CS and continuum clumps are:
$1.1-3.2$ and $1.0-4.0$, with mean values 1.5(0.1) and 1.3(0.2), respectively.
This parameter could be used for selecting nearly circular cores and exploring
radial dependences of physical parameters within the clumps.
Such an analysis lies beyond the scope of the present paper and we postpone
it to future publications.

\section{Physical parameters derived from continuum data and from LVG modeling}

Using our 1.2~mm data and IRAS fluxes it is possible to estimate
physical parameters of the clumps including dust temperatures, masses,
hydrogen column densities and luminosities.
The CS(5--4) data together with the (2--1) data (Zinchenko et al. 1995)
allowed us to calculate LVG densities, CS column densities and molecular
abundances for eight sources.

\subsection{Dust temperatures, masses, hydrogen column densities
and luminosities}
\label{sec:dustpar}

For the clumps with nearby IRAS sources we have derived dust color temperatures
by fitting the IRAS and 1.2~mm total fluxes
($F_{\rm total}$, Table~\ref{clumps2}) with a
2-temperature greybody curve and assuming optically thin conditions:
$A_1\nu^{\,\beta}B_{\nu}(T_1)+A_2\nu^{\,\beta}B_{\nu}(T_2)$
(Mozurkewich et al. 1986).
We exclude G269.11 from the analysis because the IRAS source lies outside
the half-maximum level in this source.
We set the power-law index of dust emissivity--frequency
dependence (\,$\beta$\,) equal to 2 in our calculations.
Setting $\beta=1$ leads to 10--20~\% increase in dust temperatures.
The values of dust temperatures ($T_{\rm d}$, cold component) are given
in column 2, Table~\ref{table:masses}.
The temperatures of the hot dust component lie in the range: $\sim 95-115~K$.
Dust masses have been calculated
according to the expression (e.g. Doty \& Leung 1994):

$$ M_{\rm d}=\frac{F_{\rm total}\,D^2}{k_{1.2}\,B_{1.2}(T_{\rm d})} ,$$

{\noindent where $D$, $k_{1.2}$ and $B_{1.2}(T_{\rm d})$ are the
source distance, the dust mass absorption coefficient and Plank function
at 1.2~mm, respectively.}
Ossenkopf \& Henning (1994) derived a mass absorption coefficient
of 1~cm$^2$ g$^{-1}$ at 1.3~mm wavelength appropriate for cold dust grains
covered with thick icy mantles.
We adopted this value for our calculations assuming that water ice cannot
evaporate at the derived values of $T_{\rm d}$ (e.g. van Dishoeck 2004).
Gas masses ($M_{\rm g}$), calculated assuming a gas-to-dust mass ratio,
$R_m\,=\,100$, lie in the range $\sim$\,90\,--\,6900~$M_{\odot}$.
Note that these values could be underestimated if the dust temperature
decreases outwards as expected in the case of internal heating.
Using the simple model of optically thin spherical dust shell with radial
temperature and density gradients (Pirogov \& Zinchenko 1998; see the Appendix)
and taking the fluxes at 1.2~mm and 100~$\mu$m for each source
it is possible to calculate the ratio of masses obtained in this model
and in the isothermal approximation.
It mainly depends on the ratio of outer-to-inner radii of the shell
and the density--radius power-law index, increasing when these two
parameters increase.
Taking the power-law index $p=2$ and a radii ratio of $10^{\,4}$
we find that the mass of the shell for some sources can be
up to 5 times higher than the mass calculated in the isothermal approximation.

Hydrogen column densities have been calculated from peak fluxes per beam
and dust temperatures (e.g. Motte et al. 1998):

$$N_{\rm H_2}=\frac{F_{\rm peak}}{\Omega\,m\,R_m^{-1}\,k_{1.2}\,B_{1.2}(T_{\rm d})} ,$$

{\noindent where $\Omega$ is beam solid angle, $m=2.33~amu$
is mean molecular mass.}
The $N_{\rm H_2}$ values given in column 5 of Table~\ref{table:masses}
fall in the range: (0.7\,--\,12.0)$\times$\,$10^{23}$~cm$^{-2}$.

\begin{table}[htb]
\centering
\caption[]{Physical parameters derived from continuum data}
\small
\begin{tabular}{lrrrr}
\noalign{\hrule}\noalign{\smallskip}
Source   &$T_{\rm d}$(K) & $M_{\rm g}(M_{\odot}$) & $L(L_{\odot}$)
         & $N_{\rm H_2}$(cm$^{-2}$) \\
\noalign{\smallskip}\hline\noalign{\smallskip}

G~264.28$+$1.48 & 32 & 93  & $6.6\times10^3$   & $7.0\times10^{22}$ \\
G~265.14$+$1.45 & 30 & 624  & $5.8\times10^4$ & $1.2\times10^{23}$ \\
G~268.42$-$0.85 & 35 & 360  & $4.6\times10^4$  & $3.2\times10^{23}$ \\
G~270.26$+$0.83 & 29 & 469  & $1.8\times10^4$  & $1.8\times10^{23}$ \\
G~285.26$-$0.05 (1) & 33 & 1418 & $4.6\times10^5$  & $2.0\times10^{23}$ \\
G 291.27$-$0.71 (1) & 25 & 6910 & $2.9\times10^5$  & $1.2\times10^{24}$\\
G~294.97$-$1.73 (1) & 27 & 98  & $5.9\times10^3$   & $9.6\times10^{22}$\\
G~316.77$-$0.02 & 28 & 2415 & $1.9\times10^5$ & $1.8\times10^{23}$\\
G~345.01$+$1.80  (1)& 30 & 639  & $7.4\times10^4$  & $2.0\times10^{23}$\\
G~345.41$-$0.94 & 29 & 1035 & $3.4\times10^5$  & $1.8\times10^{23}$\\
G~351.41$+$0.64 (1) & 30 & 591  & $8.5\times10^4$  & $6.5\times10^{23}$\\
\noalign{\smallskip}\hline\noalign{\smallskip}
\end{tabular}
\label{table:masses}
\end{table}

The comparison of $M_{\rm g}$ masses with CS (Table~\ref{clumps1})
and N$_2$H$^+$ virial masses (Paper~I),
for closely located clumps, shows that they agree
within a factor of 2 except for G291.27, where $M_{\rm g}$
exceeds CS virial mass by a factor of 4.5.
As the uncertainties in dust masses
(mainly due to uncertainties in distance and in dust absorption coefficient
and due to deviations from isothermal approximation) can be rather high,
we conclude that the two mass estimates are in reasonable agreement.

Bolometric luminosities ($L$) calculated by integrating the fitting curves
over total frequency range are given
in column 4 of Table~\ref{table:masses}.
They lie in the range: (0.6\,--\,46.0)$\times$\,$10^4~L_{\odot}$
indicating the presence of embedded high-mass objects.
The $L/M_{\rm g}$ ratio which is considered to be proportional
to star formation rate ranges from 38 to 325 $L_{\odot}/M_{\odot}$
for the sources in Table~\ref{table:masses}
with a mean value $139\pm 105~L_{\odot}/M_{\odot}$.
This is consistent with $\langle L/M\rangle$ estimates
for large samples of high-mass star forming regions
with masses obtained from dust continuum data:
  $71\pm 56~L_{\odot}/M_{\odot}$ (Faundez et al. 2004),
$136\pm 100~L_{\odot}/M_{\odot}$ (Shirley et al. 2003),
$120\pm 90~L_{\odot}/M_{\odot}$ (the data from Beuther et al. 2002
rescaled by Mueller et al. 2002).

\subsection{LVG densities and molecular abundances}
\label{sec:lvg}

Taking into account peak line intensities of two different CS transitions
one can derive hydrogen densities for given kinetic temperature
using the LVG approach.

Nine sources of our sample were previously mapped
in the CS(2--1) line with the SEST antenna (Zinchenko et al. 1995)
with a 50$''$ beam and 40$''$ grid spacing.
The peak CS(2--1) positions were also observed in the CO(1--0) line
by these authors.
We have compared their data with our CS(5--4) data for eight sources
(except G267.94).

In order to use both the CS(2--1) and the CS(5--4) data in model calculations
it is important to convolve them to the same beam.
For this purpose we have smoothed the CS(5--4) main beam temperature maps
to the 50$''$ resolution using our 2D Gaussian fitting program
and have calculated errors of the convolved CS(5--4) temperatures
using the propagation of errors approach.
We applied the CS--H$_2$ collisional rates from Turner et al. (1992)
which are tabulated for given kinetic temperatures.
The model is isothermal and kinetic temperatures have been estimated
according to the temperatures of the CO(1--0) line and of the dust
(cold component) (Table~\ref{table:masses}).
The assumed values of kinetic temperatures are given in Table~\ref{abundances}.
Twelve CS rotational levels have been considered in the calculations.

\begin{figure}
 \centering \includegraphics[width=8.5cm]{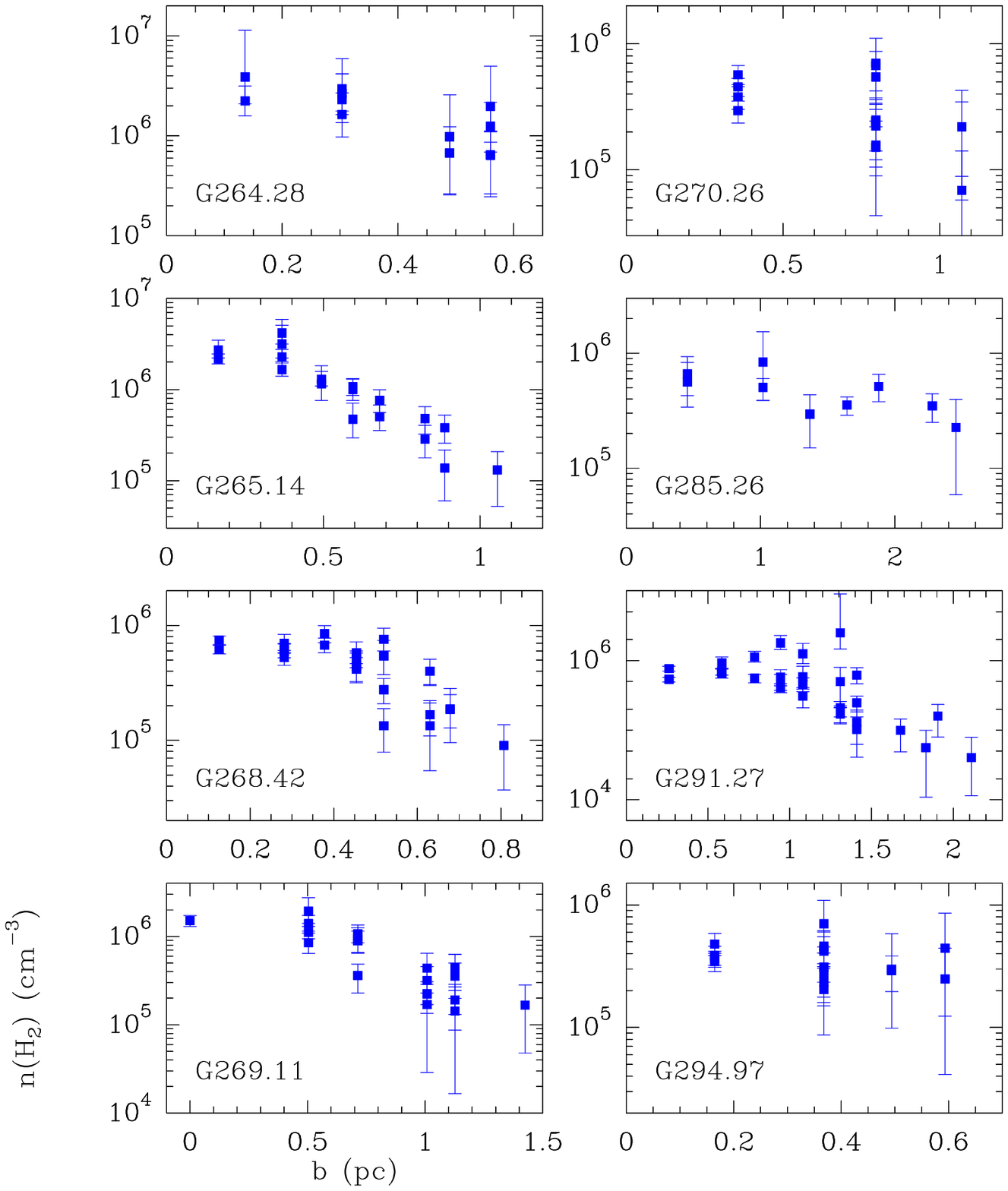}
\caption{LVG densities for eight sources
versus projected distance from the CS(5--4) peaks.
Error bars correspond to 1$\sigma$ uncertainty in the CS line temperatures}
\label{1d-dens}
\end{figure}

The results of model calculations for eight analyzed sources are shown
in Fig.~\ref{1d-dens} as density versus projected distance
from the positions of peak CS(5--4) integrated intensities (Table~\ref{table:peak}).
Central densities vary in the range:
$(3-40)\times 10^5$~cm$^{-3}$.
In all but two sources (G270.26 and G294.97) density clearly decreases
with distance from the CS(5--4) peaks.
In four sources (G265.14, G268.42, G269.11 and G291.27) density falls
by about an order of magnitude at $\sim$\,0.8--2~pc from the center.
The slopes of density--radius dependences for the outer regions
of these sources vary from --2.0 to --3.6.
In G265.14 and G291.27 these trends may be due to the
existence of neighboring low density clumps (see Fig.~\ref{maps}).
Note that the calculated LVG densities depend on the adopted kinetic
temperatures (the lower the temperature, the higher the density).
If the kinetic temperature decreases outwards, calculated densities
far from the center could be underestimated.
Changing the temperature from 40~K to 20~K leads to a density increase
by about half an order of magnitude.
Also, densities could be slightly overestimated
towards the positions of IRAS sources due to the influence
of infrared pumping on CS excitation (see Section~\ref{discussion}).

Taking into account density estimates, it is possible to calculate
gas masses for the most intensive CS(5--4) clumps in each source.
Assuming spherically-symmetric clumps with radius $d/2$, where $d$
is the CS(5--4) clump size (Table~\ref{clumps1}),
the masses could be calculated as:
$M=4\pi/3\,m\,\langle n\rangle\,(d/2)^3$, where $\langle n\rangle$
is the density averaged over positions within the area of radius $d/2$.
Using these masses we calculate mass ratios, $M_{\rm vir}/M$,
that should be considered as dense gas volume filling factor.
The mean mass ratio for eight sources is $0.2\pm 0.2$.
This value is consistent with the one found by Shirley et al. (2003)
for their sample of about 40 dense high-mass star forming regions
($0.5\pm 0.7$).
It is also consistent with the estimate $\la 0.2$ for dense gas
volume filling factor found by Juvela (1998) from multi-line CS and C$^{34}$S
profile modeling towards a sample of southern high-mass star forming regions.

Using CS column densities from LVG calculations and H$_2$ column
densities from 1.2~mm continuum data smoothed to the 50$''$ beam,
the CS abundances have been calculated for eight sources
towards positions close to IRAS sources.
The calculated values are given in Table~\ref{abundances}.
For the same positions we have calculated the N$_2$H$^+$ LTE column densities
(at $T_{\rm EX}=10$~K) and the N$_2$H$^+$ abundances.
The abundances lie in the following ranges: $X$(CS)=$(0.3-2.7)\times10^{-9}$,
$X$(N$_2$H$^+$)=$(0.3-4.4)\times10^{-10}$.
The lowest $X$(N$_2$H$^+$) values have been found towards the most luminous IRAS sources.
Mean N$_2$H$^+$ abundances for several clumps were calculated in Paper~I
using N$_2$H$^+$ virial mass estimates.
There is a reasonable agreement between different $X$(N$_2$H$^+$)
estimates for clumps in G265.14, G269.11 and G294.97.
For G268.42 our estimate of $X$(N$_2$H$^+$) is about 4 times lower
than the estimate from Paper~I.
For G285.26 and G291.27 our $X$(N$_2$H$^+$) values are about an order
of magnitude lower than the estimates from Paper~I, but they relate
to different clumps, implying abundance variations over the sources.
The role of abundance variations is discussed in Section~\ref{discussion}.

\begin{table}[htb]
\centering
\caption[]{Molecular column densities and abundances smoothed over
50$''$ Gaussian beam towards positions close to IRAS sources}
\scriptsize
\begin{tabular}{lrrrrrc}
\noalign{\hrule}\noalign{\smallskip}
Source,  & $T_{\rm KIN}$ & $N$(H$_2$)  & $N$(CS)  & $X$(CS)     & $N$(N$_2$H$^+$) & $X$(N$_2$H$^+$)\\
position &  (K)          & (cm$^{-2}$)   & (cm$^{-2}$)  & ($10^{-9}$) & (cm$^{-2}$)     & ($10^{-10}$)   \\
         &               & ($10^{22}$)   & ($10^{13}$)  &             & ($10^{12}$) \\
\noalign{\smallskip}\hline\noalign{\smallskip}

G~264.28 (0,--40) & 20 & 3.1  & 3.4  & 1.1 & 4.1  & 1.3 \\
G~265.14 (0,--20) & 20 & 4.7  & 4.8  & 1.0 & 16.1 & 3.4 \\
G~268.42 (0,0)    & 40 & 13.8 & 14.0 & 1.0 & 8.7  & 0.6 \\
G~269.11 (0,0)    & 20 & 3.4  & 4.5  & 1.3 & 14.0 & 4.2 \\
G~270.26 (0,0)    & 40 & 2.9  & 7.9  & 2.7 & 13.0 & 4.4 \\
G~285.26 (0,0)    & 40 & 6.0  & 4.4  & 0.7 & 1.6  & 0.3 \\
G 291.27 (20,--20)& 40 & 54.5 & 17.5 & 0.3 & 15.6 & 0.3 \\
G~294.97 (0,0)    & 40 & 4.4  & 5.2  & 1.2 & 11.0 & 2.5 \\
\noalign{\smallskip}\hline\noalign{\smallskip}
\end{tabular}
\label{abundances}
\end{table}

Note, that other physical parameters found for the sample cores (sizes,
aspect ratios, densities, masses, dust temperatures and luminosities)
lie within the ranges found in the large surveys of high-mass star forming
regions performed during recent years in CS(5--4) (Plume et al. 1997,
Shirley et al. 2003) and 1.2~mm continuum (Beuther et al. 2002,
Mueller et al. 2002, Faundez et al. 2004, Fontani et al. 2005).

\section{Comparison of molecular and continuum data}
\label{sec:comparison}

The maps in Fig.~\ref{maps} show in many cases significant
differences between CS(5--4) and N$_2$H$^+$(1--0) intensity distributions
as well as between 1.2~mm continuum and N$_2$H$^+$(1--0).
We have performed detailed point-to-point comparison of molecular intensities
and continuum fluxes for each source and the results are given below.
In addition, we have compared the CS(5--4) and the N$_2$H$^+$(1--0) line velocities
and line widths.

\subsection{CS and dust maps}

The comparison of the CS(5--4) and 1.2~mm continuum data shows
that in most cases the CS intensities depend almost linearly
on continuum fluxes implying that CS(5--4) is a good tracer
of total gas column density.
Some deviations from linear dependences could be connected
with differences in excitation conditions in clumps within the same source.

\subsection{CS and N$_2$H$^+$ maps}

In many cases (especially in G285.26, G291.27, G345.01, G345.41
and G351.41, see Fig.~\ref{maps}) the N$_2$H$^+$(1--0) and the CS(5--4)
intensities have different spatial distributions which, in general,
could be caused by opacity effects, line excitation or differences in chemical
composition of individual clumps.

\begin{figure*}
 \centering \includegraphics[width=15cm]{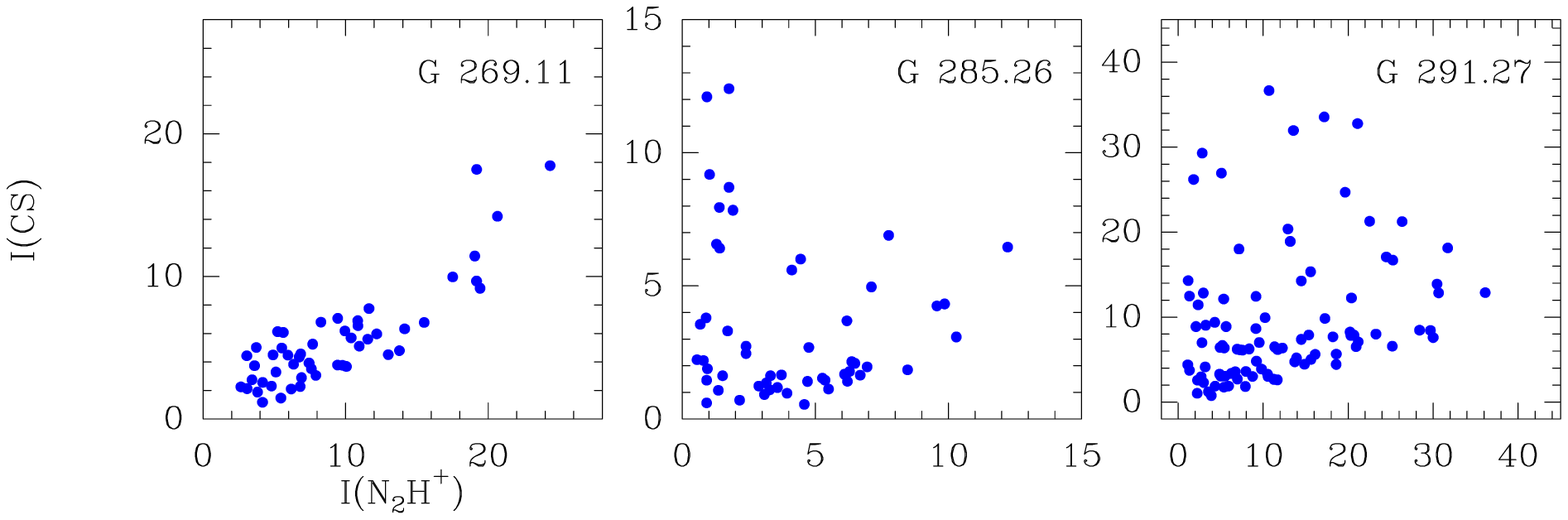}
\caption{The CS(5--4) versus N$_2$H$^+$(1--0) integrated intensities
(both in K~km~s$^{-1}$) for three representative sources.
The CS data is smoothed to the 50$''$ resolution}
\label{ics-in2h}
\end{figure*}

For comparison with the N$_2$H$^+$(1--0) data obtained
with the $55''$ beam, the CS(5--4) integrated intensities have been averaged
over areas with 25$''$ radius around each position.
In five objects (G265.14, G268.42, G269.11, G270.26 and G294.97)
$I$(CS) almost linearly depends on $I$(N$_2$H$^+$).
Other sources show a low correlation or no correlation.
In some of them (G285.26, G345.01 and G351.41) the plots split
into separate branches which probably belong to different clumps.
In Fig.~\ref{ics-in2h} the CS(5--4) integrated intensities
are shown versus $I$(N$_2$H$^+$) for three representative sources
with high correlation (G269.11) and low correlation (G285.26 and G291.27)
between intensities.

Comparison of the N$_2$H$^+$(1--0) maps with the CS(2--1) maps
(Zinchenko et al. 1995) shows that in some cases (G264.28, G285.26 and G291.27)
they are in better agreement than with the CS(5--4) maps.

\subsection{Intensity ratios}
\label{iratios}

In order to study spatial variations of molecular and continuum emission
within the sources we have calculated the CS(5--4) to the N$_2$H$^+$(1--0)
integrated intensity ratios as well as the ratios of molecular integrated
intensities to continuum fluxes ($I$(CS)/$F_{1.2}$ and
$I$(N$_2$H$^+$)/$F_{1.2}$) for each source position.
The continuum fluxes ($F_{1.2}$) have been averaged over areas with 25$''$
radius around each N$_2$H$^+$ position for $I$(N$_2$H$^+$)/$F_{1.2}$ dependences.
The ratios of molecular integrated intensities to continuum fluxes
give information on abundance variations over the sources,
if lines are optically thin as in the case of N$_2$H$^+$(1--0) (Paper~I).
We used data with signals higher than 3$\sigma$, and rejected map
positions with intensities below 5\% the peak values (3\% for G268.42
and G291.27), to avoid spikes of ratios which sometimes occur
at the edges of the maps.

\begin{figure*}
 \centering \includegraphics[width=15cm]{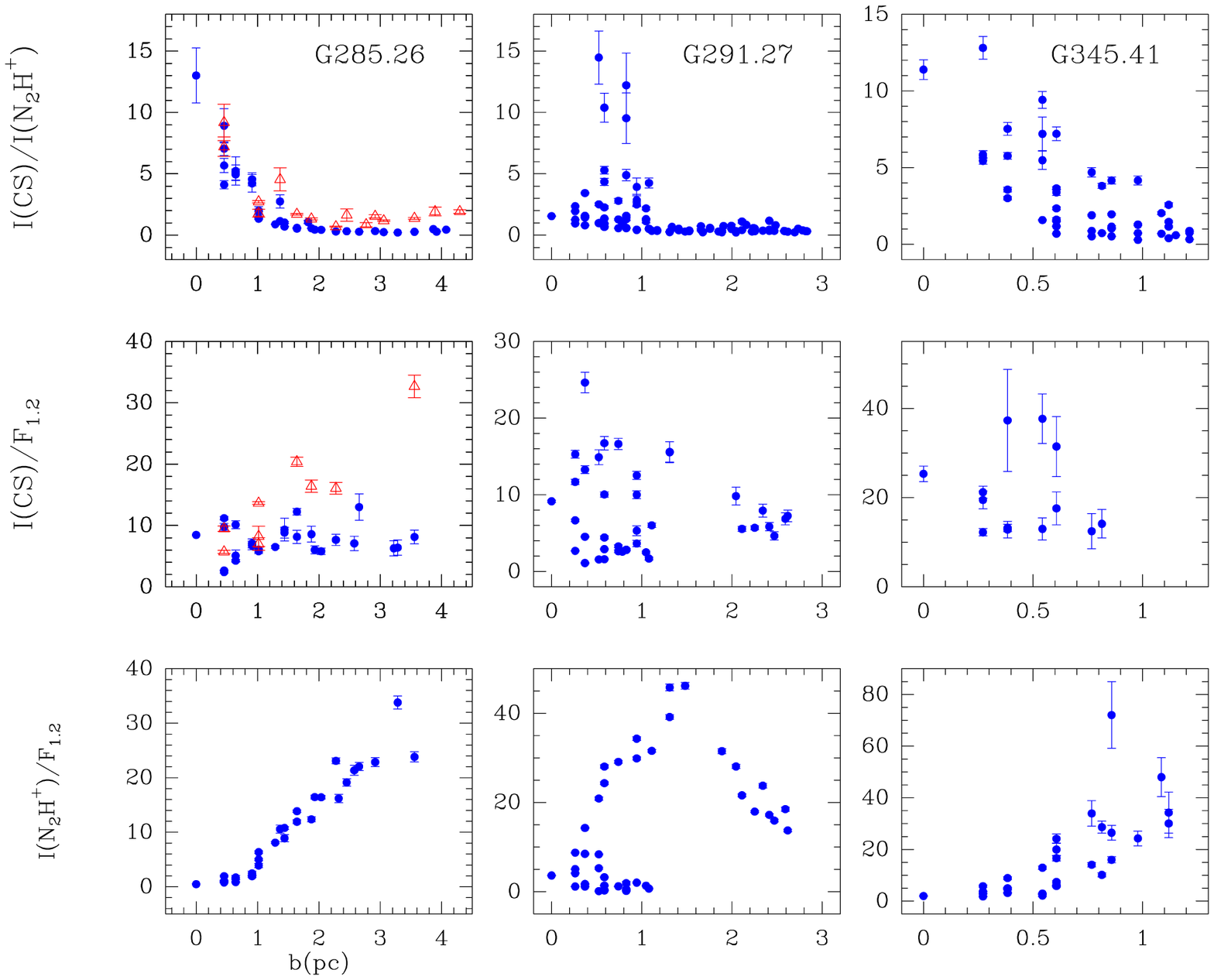}
\caption{The CS(5--4)/N$_2$H$^+$(1--0)
integrated intensity ratios and the ratios of molecular intensities
to continuum fluxes versus projected distance from the CS(5--4) peaks
for three sources with most prominent differences between CS and N$_2$H$^+$
(in lower panels, $F_{1.2}$ is the flux at
1.2~mm in Jy~beam$^{-1}$ averaged over areas with 25$''$
radius around each N$_2$H$^+$ position).
In addition, for G285.26 the CS(2--1)/N$_2$H$^+$(1--0) and
the CS(2--1)/continuum intensity ratios are shown by red triangles.
In this case continuum fluxes have been averaged over areas
with 25$''$ radii around each CS(2--1) position.
Error bars correspond to 1$\sigma$ uncertainties of the ratios}
\label{irat3}
\end{figure*}

The dependences of $I$(CS)/$I$(N$_2$H$^+$) (upper panels),
$I$(CS)/$F_{1.2}$ (middle panels) and $I$(N$_2$H$^+$)/$F_{1.2}$ (lower panels)
with respect to projected distance from the CS(5--4) peaks ($b$) are shown
in Fig.~\ref{irat3} for three selected sources,
where differences between the CS(5--4) and
N$_2$H$^+$(1--0) intensity distributions are among the most prominent.
In several sources (especially in G285.26, G345.41 and G351.41)
the $I$(CS)/$I$(N$_2$H$^+$) ratios have a tendency to decrease with distance.
Six sources demonstrate variations of $I$(CS)/$I$(N$_2$H$^+$)
across the source by more than an order of magnitude
(G285.26, G291.27, G345.01, G345.41 and G351.41).
The highest ratios ($\sim$\,13\,--\,15) are observed in G285.26, G291.27
and G345.41, in the vicinity of the most luminous IRAS sources of our sample.
The lowest ratios ($\la 0.1$) are observed in G316.77, G345.01 and G351.41
towards positions close to the centers
of the N$_2$H$^+$ clumps without CS counterparts and IRAS sources.

There is a rather large scatter
both in $I$(CS)/$F_{1.2}$ and $I$(N$_2$H$^+$)/$F_{1.2}$ ratios
across the sources (about an order of magnitude).
The $I$(N$_2$H$^+$)/$F_{1.2}$ ratios clearly drop towards CS peaks
in most cases while the $I$(CS)/$F_{1.2}$ ratios demonstrate no such trends.
No clear trends of $I$(N$_2$H$^+$)/$F_{1.2}$ versus $b$ have been found
in G268.42 and G294.97.

For several sources we have calculated ratios of the CS(2--1)
both to the N$_2$H$^+$(1--0) integrated intensities and to continuum fluxes
averaged over areas with 25$''$ radii around each CS(2--1) position.
It is found that these ratios show similar trends with $b$
as the $I$(CS)/$I$(N$_2$H$^+$) ratios for CS(5--4)
and the $I$(N$_2$H$^+$)/$F_{1.2}$ ratios, respectively
(see plots for G285.26 in Fig.~\ref{irat3}).
For CS(2--1) an increase of the $I$(CS)/$F_{1.2}$ ratios with $b$
is probably connected with line saturation
(see Section~\ref{discussion} and Appendix~\ref{modeling}).

The comparison of intensity ratios at positions close to IRAS sources
with bolometric luminosities (Table~\ref{table:masses}) shows
that the $I$(CS)/$I$(N$_2$H$^+$) ratios
are clearly enhanced towards the three most luminous sources (G285.26, G291.27
and G345.41), the $I$(N$_2$H$^+$)/$F_{1.2}$ ratios drop towards these sources.

\subsection{CS and N$_2$H$^+$ velocities and line widths}
\label{vanalysis}

The velocity comparison of two molecular lines with different optical depths
gives information about internal kinematics of the sources.
Mardones et al. (1997) have introduced normalized velocity difference:
$\delta V=(V_{thick}-V_{thin})/\Delta V_{thin}$,
where $V_{thick}$ and $V_{thin}$ are optically thick and thin
line velocities, respectively, $\Delta V_{thin}$ -- optically thin line width.
Their analysis of molecular lines in low-mass cores has proved
this parameter to be an effective tracer of infall
(Mardones et al. 1997, Lee et al. 1999, 2001) and/or core accretion motions
(e.g. Tafalla et al. 2002).
Although N$_2$H$^+$ and CS are likely to trace different regions,
as in low mass cores, it is still interesting to quantify the relative
motions between different parts of the same cloud and see if there are
signs of systematic contraction or expansion.

The N$_2$H$^+$ lines most probably are optically thin in the sources
(Paper~I) while the CS lines could be optically thick.
In order to search for systematic motions we have calculated normalized
velocity differences between CS and N$_2$H$^+$ ($\delta V_{\rm CS}$)
taken from Gaussian fits for each source position.
Both the CS(5--4) and CS(2--1) data (Zinchenko et al. 1995) have been used.
Given that the signal-to-noise ratios of the CS(5--4) data are rather low,
we have only analyzed the spectra with integrated intensities not lower
than 20\%--30\% of the maximum value.
Even if our spatial resolution does not permit to resolve the regions
with systematic motions, one could expect to detect non-zero normalized
velocity differences towards positions close to density peaks.

According to the criterion $|\delta V|>0.25$ from Mardones et al. (1997)
no indications of systematic motions have been found towards CS(5--4) peaks.
However, we have found such indications towards some other positions
in several sources.
In particular, in G285.26 the $\delta V_{\rm CS}$ values
towards the (0'',0'') position are $-$0.58(0.15) and $-$0.73(0.19)
for CS(5--4) and CS(2--1), respectively, probably implying infall motions.

Both CS and N$_2$H$^+$ line widths are much higher than the
thermal widths, and the line profiles in most sources are nearly symmetric
without prominent self-absorption features.
The CS(2--1) lines are broader than the N$_2$H$^+$(1--0) lines
in most cases (mean line width ratios vary from source to source
in the range: 0.9--1.7).
The CS(5--4) lines are narrower than CS(2--1) and their widths are on average
close to the N$_2$H$^+$(1--0) widths implying
that the CS(5--4) optical depths are not high
(mean line width ratios vary in the range: 0.7--1.3).
In some sources (G269.11, G285.26, G291.27, G345.01 and G345.41)
the CS(5--4) line widths are broader towards CS peaks
and decrease outwards.
Similar trends found in Paper~I for the N$_2$H$^+$(1--0) line widths
were explained as an influence of enhanced dynamical activity
in the central regions of the cores in the vicinity of IRAS sources.
In the case of CS(5--4), these trends in addition could be connected
with enhanced optical depth towards CS peaks.

The clumps where N$_2$H$^+$ drops towards CS peaks (Section~\ref{iratios})
have enhanced CS(5--4) mean line widths ($\ga 4$~km s$^{-1}$,
Table~\ref{clumps1}) implying either enhanced optical depth or
some dynamical processes which broaden line profiles towards these peaks.
There is also a weak correlation between $\Delta V$(CS) and luminosities
for positions close to IRAS sources.

\section{Discussion}
\label{discussion}

Here we examine possible reasons for the observed variations
of the CS/N$_2$H$^+$, CS/dust and N$_2$H$^+$/dust ratios.

{\bf Opacity effects}.
The effect of line saturation due to opacity leads to a decrease of the
relative line intensities towards CS and dust emission peaks (where column density
is presumably the highest one).
We have not seen such effect for the CS(5--4) data and this can be considered
as an implicit argument against high optical depth in these lines.
On the other hand a decrease of CS(2--1) relative intensities towards density
peaks in several sources could be connected with saturation.
Using the C$^{34}$(2--1) data for several sample sources (Zinchenko et al. 1995)
we have found that the optical depth of the CS(2--1) line is higher than unity
implying a decrease of the CS(2--1) intensity compared with the
optically thin case.
For G285.26~($-40'',-40''$), $\tau=2.5$ and the CS(2--1) intensity decreases
2.7 times.
The N$_2$H$^+$(1--0) lines most probably are not optically thick (Paper~I)
within the 55$''$ beam and it is hard to consider opacity effects
as an explanation of the observed variations on the scales resolved by the beam.

{\bf Excitation}.
Molecular line excitation is determined by collisions
and radiative transfer processes.
Differences in critical densities of the CS(5--4) and N$_2$H$^+$(1--0) lines
by about an order of magnitude
can produce differences in their excitation and intensity distributions.
The analysis given in Appendix~\ref{modeling} shows that
density, kinetic and dust temperature gradients
can cause a decrease of the $I$(N$_2$H$^+$)/$F_{1.2}$ ratios
by a factor not more than 3 in the density range $10^5$\,--\,$10^6$~cm$^{-3}$
when $T_{\rm KIN}$ and $T_{\rm d}$ varies from 20~K to 40~K.
Yet, they cannot be the reason for the observed drops of these ratios
by an order of magnitude or higher, as, for example, in G285.26 where
high density contrasts are not observed (Fig.~\ref{1d-dens})
and in other sample sources.

In principle, the observed variations could be connected with radiative
transfer effects caused by pumping from nearby luminous infrared
sources which can overpopulate upper rotational levels of the molecule
and underpopulate lower levels.
This effect in our case could make the CS(5--4) line intensities higher
and the N$_2$H$^+$(1--0) line intensities lower compared with the case of
no pumping.
However, Carroll \& Goldsmith (1981) showed that for CS the size of
the region affected by IR pumping via vibrational transitions is $\la 0.03$~pc.
We have done similar analysis for N$_2$H$^+$ using parameters
of vibrational transitions from Botschwina (1984), Owrutsky et al. (1986)
and Heninger et al. (2003).
It is found that the size of the affected region considering the bending mode
is $\la 0.02$~pc and is several times lower if one considers pumping via
transitions in the stretching modes.
The details of calculations are given in the Appendix~\ref{pumping}.
A possible radiative pumping via rotational transitions due to the background
radiation field, including IR source, is also found to be unimportant
for our sources (Appendix~\ref{pumping}).

{\bf Chemistry}.
As both the opacity and molecular line excitation effects are found
to be insufficient, the observed variations of molecular intensity
to continuum flux ratios could mainly be connected with abundance variations
of these molecules.
Moreover, for different clumps within G285.26 and G291.27
the N$_2$H$^+$ abundances are found to vary by an order of magnitude
(see Section~\ref{sec:lvg}).
High CS-to-N$_2$H$^+$ intensity ratios previously have been found
towards some starless clumps, for example, in S68N  (Williams \& Myers, 1999)
and in Perseus (Olmi et al. 2005).
They can be explained by N$_2$H$^+$ underabundance
which occurs at early phases of core evolution (Bergin et al. 1997).
However, in our cores the high CS-to-N$_2$H$^+$ intensity ratios
are observed in the vicinity of IRAS sources indicating that they
are evolved regions.
The drop of the N$_2$H$^+$ intensity towards YSOs previously was observed
in Orion (Ungerechts et al. 1997) and Cepheus A (Bottinelli \& Williams 2004).
Hot core chemical models (e.g. Nomura \& Millar 2004) where
ultraviolet radiation from embedded stars leads to evaporation
of grain mantles and CS enhancement, cannot explain why N$_2$H$^+$
intensities do not correlate with gas column density in these regions.

Recently Lintott et al. (2005) have proposed an alternative model
where CS to N$_2$H$^+$ abundance ratio becomes enhanced due to higher
collapse rates compared to free-fall regime (accelerated collapse).
As a result, high densities are achieved before CS and the molecules
responsible for N$_2$H$^+$ removal are freezing onto grains.
Yet, due to gradients of excitation temperatures,
systematic collapse motions predicted by this model probably
should give asymmetric or self-absorbed profiles of optically thick lines
(e.g. CS(2--1)) which are not observed in most of the sources.
Normalized velocity differences (Section~\ref{vanalysis})
have not indicated such motions towards CS peaks
in our sources according to the Mardones et al. (1997) criterion.
Yet, if the sources consist of a large number of small unresolved clumps
with low volume filling factor as implied in Section~\ref{sec:lvg},
the observed CS lines could have low effective
optical depth and nearly symmetric profiles.
In this case systematic collapse motions could be detected as additional
line broadening towards core centers.
The enhanced CS(5--4) line widths have been detected towards
intensity peaks in several sources (Section~\ref{vanalysis})
where the $I$(N$_2$H$^+$)/$F_{1.2}$ ratios are low.
If the CS lines actually are not optically thick, these enhancements
could imply some dynamical processes (e.g. collapse motions)
in gas traced by CS.
In order to decide whether or not the accelerated collapse model
could be applied to these cores, one needs to conduct detailed
radiative transfer calculations in comparison with observational data
including other molecular tracers (e.g. HCO$^+$, NH$_3$, SO).

Another possible explanation of the N$_2$H$^+$ abundance drop
can be related with its dissociative recombination.
Earlier it was assumed that dissociative recombination of N$_2$H$^+$ leads
primarily to the formation of molecular nitrogen which can form
N$_2$H$^+$ again via reaction with $\mathrm{H_3^+}$.
However, Geppert et al. (2004) found that another channel
which leads to formation of the NH and H molecules dominates.
Perhaps this could be the reason why the N$_2$H$^+$ abundance drops
in vicinities of luminous YSOs.
It is well known that UV radiation penetrates deep into dense molecular
clouds, probably due to their clumpy structure (e.g. Meixner et al. 1992,
Meixner \& Tielens 1993).
Therefore, one can expect an enhanced ionization in the neighborhood
of luminous young stars with a high UV flux, on the scales comparable
to those which are discussed here.
However, published results on ionization fraction in massive cores
do not support this view (Bergin et al. 1999).
On the other hand, preliminary estimates of this fraction from
our measurements for similar sources in the northern sky do
show noticeable variations within the sources
(Zinchenko et al, in preparation).
A conclusive decision on the role of this factor in the observed chemical
differentiation needs further investigation including chemical modeling.

\section{Conclusions}

In order to get reliable information on the density and chemical structure
of the dense cores associated with high-mass star forming regions
twelve objects from southern hemisphere have been mapped in the CS(5--4)
line and in dust continuum at 1.2~mm.
We have compared CS(5--4) and continuum data with each other as well as
with the N$_2$H$^+$(1--0) (Paper~I) and the CS(2--1) (Zinchenko et al. 1995) data.
Besides, physical parameters of the cores have been derived
from the CS(5--4) and the continuum data.

The results can be summarized as follows:

\begin{enumerate}
\item
Most of the maps have several emission peaks (clumps).
Mean sizes of 17 clumps, having counterparts in continuum and CS,
are 0.30(0.06)~pc (continuum) and 0.51(0.07)~pc (CS).
The CS virial masses lie in the range $\sim$140\,--\,1630~$M_{\odot}$.
For the clumps with IRAS sources we derived dust temperatures: 24\,--\,35~K,
masses: 90--6900~$M_{\odot}$, molecular
hydrogen column densities: (0.7\,--\,12.0)$\times$\,$10^{23}$~cm$^{-2}$
and luminosities: (0.6\,--\,46.0)$\times$\,$10^4~L_{\odot}$.

\item
Using the CS(5--4) and the CS(2--1) data (Zinchenko et al. 1995)
LVG densities in eight sources have been calculated.
Densities towards CS peaks within the 50$''$ beam (0.56~pc at 2.3~kpc,
the average distance of our sample source) vary from source to source
in the range: (3\,--\,40)$\times$\,$10^5$~cm$^{-3}$.
In four sources, the density falls by about an order of magnitude
at $\sim$\,0.8\,--\,2~pc from the CS peaks
which could be connected either with the existence of lower density clumps
or density gradients within clumps.
The masses calculated from LVG densities are higher
than virial masses and masses derived from continuum data
implying small-scale clumpiness of the cores.
The CS abundances have been calculated towards IRAS positions
within 50$''$ Gaussian beam for eight sources:
$X$(CS)=(0.3\,--\,2.7)$\times$\,$10^{-9}$.
The N$_2$H$^+$ abundances for the same positions lie in the range:
$X$(N$_2$H$^+$)=(0.3\,--\,4.4)$\times$\,$10^{-10}$.

\item
For most of the objects the CS and continuum peaks are close
to the IRAS point source positions.
The CS(5--4) integrated intensities correlate with continuum fluxes per beam
almost linearly in all cases while significant correlations between
the CS(5--4) and the N$_2$H$^+$(1--0) integrated intensities are found
only in five sources.
The highest CS to N$_2$H$^+$ integrated intensity ratios
($\sim 10$) are found in the vicinity of the most luminous IRAS sources.
The lowest ratios ($\la 0.1$) are observed
towards positions close to the centers
of the N$_2$H$^+$ clumps without CS counterparts and IRAS sources.

\item
The study of spatial variations of molecular integrated intensity ratios
to continuum fluxes reveals that
for most of the sources the $I$(N$_2$H$^+$)/$F_{1.2}$ ratios drop
towards CS and dust emission peaks near IRAS sources, likely
implying underabundance of N$_2$H$^+$ towards these peaks.
The $I$(CS)/$I$(N$_2$H$^+$) ratios are clearly enhanced towards the three most
luminous sources, whereas the $I$(N$_2$H$^+$)/$F_{1.2}$ ratios drop
towards these sources.
Possible explanations of these results could involve dissociative recombination
of N$_2$H$^+$ or accelerated collapse processes.
For CS(5--4) the $I$(CS)/$F_{1.2}$ ratios show no clear trends with distance
from the CS peaks while for CS(2--1) such ratios drop towards these peaks
implying line saturation effects in several sources.

\item
The analysis of normalized velocity differences between CS and N$_2$H$^+$ lines
has not revealed indications of systematic motions in the sources
towards CS peaks.
A possible indication for infall motions has been found towards
one position in G285.26.
The CS(5--4) line widths in several sources have a tendency to decrease
with distance from CS peaks which could be connected both with optical depth
effects and with enhanced dynamical activity towards centers of the cores.

\end{enumerate}

\begin{acknowledgements}

We are grateful to Alexander Lapinov for helpful discussions and suggestions
and to Malcolm Walmsley for important comments.
We would like to thank Marie-Lise Dubernet for sending us tables with
the N$_2$H$^+$--He collisional rates and Luca Dore for providing us reprints
of the papers on the parameters of N$_2$H$^+$ vibrational transitions.
We would also like to thank the referee, Jorma Harju, for thorough examining
the manuscript, useful comments and recommendations
which significantly improved the paper.
The research has made use of the SIMBAD database,
operated by CDS, Strasbourg, France.
The work was supported by INTAS grant 99-1667 and Russian Foundation
for Basic Research grants 03-02-16307 and 06-02-16317
and by the Program ``Extended objects in the Universe"
of the Russian Academy of Sciences.

\end{acknowledgements}

{}

\appendix

\section{The CS and N$_2$H$^+$ LVG modeling results}
\label{modeling}

\begin{figure}
 \centering \includegraphics[width=8cm]{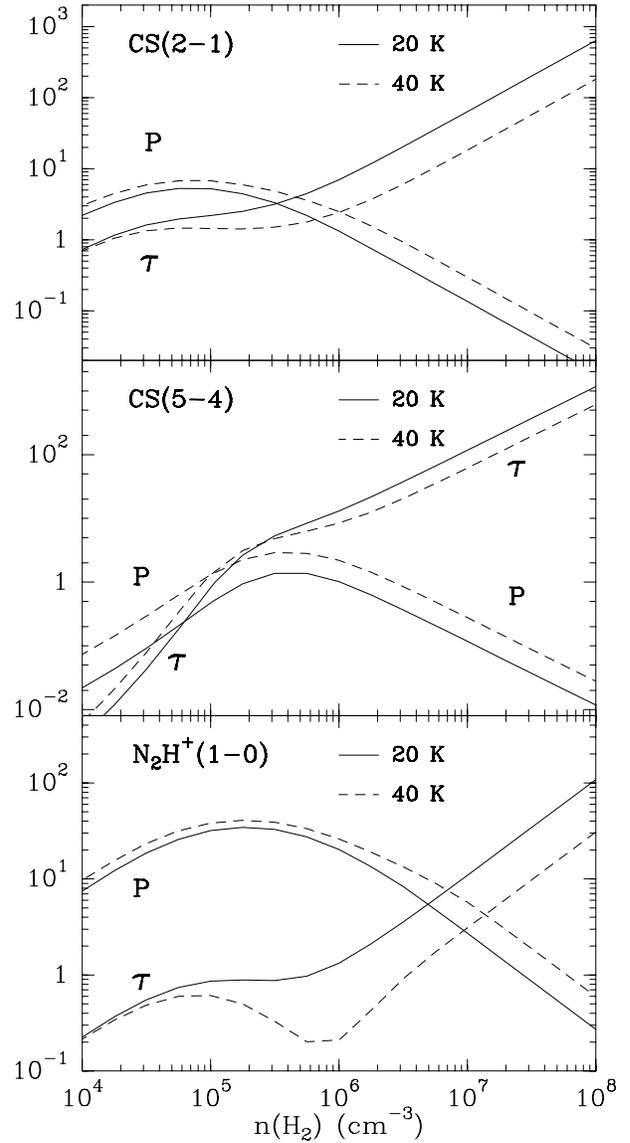}
\caption{The results of LVG modeling: CS(2--1), CS(5--4) and
N$_2$H$^+$(1--0) brightness temperatures multiplied by the factor
$10^{14} (N/V)^{-1}$ ({\bf $P$}) versus density
at two values of kinetic temperature.
The dependences of line optical depths ({\bf $\tau$}) versus density
are also shown
}
\label{lvg}
\end{figure}

Using the LVG model we have examined the CS and N$_2$H$^+$ excitation
for a wide density range at distinct kinetic temperatures.
When calculating the excitation of the N$_2$H$^+$ unsplitted rotation
lines, we used the N$_2$H$^+$--He collisional rates from Daniel et al. (2005)
(http://www.obspm.fr/basecol).
In Fig.~\ref{lvg} the CS(2--1), CS(5--4) and N$_2$H$^+$(1--0)
brightness temperatures divided by $N/V$ (the ratio of gas column density
to velocity at the boundary) are shown versus density
at two values of kinetic temperature: 20~K and 40~K.
For convenience of presentation they are multiplied by
the factor of 10$^{14}$ and are denoted as $P$.
They can be compared with observed ratios of molecular line intensities
to continuum fluxes (the latter are proportional to gas column density).
The dependences of line optical depths ($\tau$) on density
are also shown.
We set $d$\,=\,0.5~pc and $\Delta V$\,=\,2.5~km/s
as typical clump size and line width (Paper~I)
and $X$(N$_2$H$^+$)\,=\,$5\times 10^{-11}$
and $X$(CS)\,=\,10$^{-9}$ (which are close to the values for G285.26
from Table~\ref{abundances}).
These parameters enter the model as $N/V$\,=\,$n$(H$_2$)\,(2\,$d\,X)/\Delta V$.

The qualitative behavior of $P$ versus $n$(H$_2$)
is similar for all the lines, yet, for CS(5--4) $P$ reaches the
maximum at higher densities than for CS(2--1) and N$_2$H$^+$(1--0).
The position of this maximum depends on $X$(CS)
(the lower $X$(CS) the higher is the density at which $P$ reaches maximum).
After the maximum, $P$ decreases with density, approaching linear dependences
(in logarithmic scale) due to line saturation effect.
As the CS(5--4) lines in our observations are not saturated,
we consider only densities $\la 10^6$~cm$^{-3}$.
Within the density range $10^5$\,--\,$10^6$~cm$^{-3}$ $P$ falls
by a factor of $\la$\,2 for N$_2$H$^+$(1--0) and by a factor
$\sim$\,2.5--4 for CS(2--1) (Fig.~\ref{lvg}, lower and upper panels).
For the CS(2--1) lines such drop could be connected with saturation.

We have no data on spatial distributions of kinetic temperatures
in our sources, yet, one-point CO(1--0) (Zinchenko et al. 1995)
and IRAS data probably do not imply temperatures significantly
higher than 40~K towards CS and dust peaks
for the regions with angular sizes $\sim$\,1$'$.
If kinetic temperature rises from 20~K to 40~K towards density peaks,
$P$ decreases slower with the resulting decreasing factor of 1.3.
Note that dust temperature gradients from 20~K to 40~K can reduce
the observed $I$(N$_2$H$^+$)/$F$ ratios by 2.3 times.
Thus, one could expect that within the density range $10^5$\,--\,$10^6$~cm$^{-3}$
the $I$(N$_2$H$^+$)/$F$ ratios could drop by a factor not more than 3
as $T_{\rm KIN}$ and $T_{\rm d}$ rise from 20~K to 40~K.

\section{Infrared pumping and excitation of N$_2$H$^+$}
\label{pumping}

Carroll \& Goldsmith (1981) derived the following criterion
for infrared pumping to have an influence on molecular rotation spectra:

$$
f\exp{\bigl (-\frac{h\nu}{kT_s} \bigr )}\ge\frac{A_{\rm rot}}{A_{\rm vibr}}~,
$$

\noindent{where $f$ is a dilution factor of the pumping source
which radiates at the temperature $T_s$, $A_{\rm rot}$ and $A_{\rm vibr}$
are the Einstein spontaneous rates for rotational and vibrational
transitions, respectively, $\nu$ is the frequency of the vibrational transition,
$h$ and $k$ is the Plank and Boltzmann constants, respectively.}
Taking $f$\,$\approx$\,$R^2/4r^2$, where $R$ is the radius of the dust source,
$r$ is the distance from the source,
and $T_s({\rm K})$\,$\approx$\,$51\,(2\times 10^{17}{\rm cm})^{1/3}\,R^{-1/3}$
(Scoville \& Kwan, 1976) one obtains:

$$
r\le \frac{R}{2}\,\sqrt{\frac{A_{\rm vibr}}{A_{\rm rot}}}\,
\exp\Bigl (-\frac{h\nu}{k}\,\frac{1}{102}\,
\frac{R^{1/3}}{(2\times 10^{17})^{1/3}} \Bigr )~.
$$

{\noindent The right-hand of this expression peaks at some $R_{\rm MAX}$,
corresponding to:}

$$
r_{\rm MAX}({\rm cm})\approx 2.87\times 10^{24}\,\exp(-3)\,
\bigl (\frac{h\nu}{k} \bigr )^{-3}\,
\sqrt{\frac{A_{\rm vibr}}{A_{\rm rot}}}~,
$$

{\noindent so that the pumping could be effective at distances not higher
than this value.}
For the N$_2$H$^+$ molecule with $\mu$\,=\,3.4\,D and
$\nu_{\rm rot}$(1--0)\,=\,93.174~GHz (Pickett et al. 1998)
$A_{\rm rot}$\,$\approx$\,$3.6\times 10^{-5}$~s$^{-1}$.
For the first vibrational stretching mode with $\nu_1$\,=\,3234~cm$^{-1}$
and $A_{\rm vibr}$\,=\,857~s$^{-1}$ (Botshwina, 1984)
$r_{\rm MAX}$\,$\approx$\,$6.9\times 10^{15}$~cm\,
$\approx$\,$2\times 10^{-3}$~pc.
For the second bending mode with $\nu_2$\,=\,698.6~cm$^{-1}$
(Owrutsky et al. 1986),
the lifetime of the excited state is $\tau$\,=\,123.1~ms (Heninger et al. 2003)
corresponding to $A_{\rm vibr}$\,=\,8.1~s$^{-1}$ and
leading to $r_{\rm MAX}$\,$\approx$\,$6.6\times 10^{16}$~cm\,
$\approx$\,$2\times 10^{-2}$~pc.
For the third stretching mode with $\nu_3$\,=\,2254~cm$^{-1}$ (Botshwina 1984),
$\tau$\,=\,530.1~ms (Heninger et al. 2003) corresponding to
$A_{\rm vibr}$\,=\,1.9~s$^{-1}$ and leading to
$r_{\rm MAX}$\,$\approx$\,$9.5\times 10^{14}$~cm\,
$\approx$\,$3\times 10^{-4}$~pc.

A possible radiative pumping via rotational transitions due to the background
radiation field, including IR source, is effective only
when the source optical depth at these frequencies is higher than unity.
From our continuum data it is easy to estimate the peak optical
depth at 1.2~mm averaged over the telescope beam in the optically thin case,
$\tau$\,=\,$F\,\Omega^{-1}\,B_{\nu}(T)^{-1}$,
where $F$ is the peak flux per beam, $\Omega$ is the beam solid angle,
$B_{\nu}(T)$ is the Plank function at frequency $\nu$ and temperature $T$.
Thus, for the strongest continuum source of our sample, G291.27,
with peak flux $F$\,=\,18.8~Jy~beam$^{-1}$ and $T$\,=\,25~K,
$\tau$\,$\approx$\,0.05 averaged over a region with a linear size
$d$\,$\sim$\,0.3~pc.
Taking a power-law index of the column density--radius dependence
of $-$1 which corresponds to the density--radius power-law index of $-$2,
one obtains that in G291.27 the size of the region, where mean optical depth
at 1.2~mm exceeds unity, is $d$\,$\sim$\,0.015~pc.
Without averaging, the size of the optically thick region should be lower.
Our calculations of CS and N$_2$H$^+$ excitation assuming that
infrared source radiates as blackbody with temperature
$T_{\rm IR}=100$~K (close on average to the hot dust temperature component
of the sample sources, see Section~\ref{sec:dustpar}),
show that the dilution factor for the pumping source should be $f\ga 10^{-3}$
in order to produce noticeable effect on the level populations.
Therefore, the radius of affected region in G291.27 at 1.2~mm
is $r=R/(2\sqrt{f})=d/(4\sqrt{f})\la 0.1$~pc.
The frequency of the N$_2$H$^+$(2--1) transition which mainly
affects the $J$\,=\,1 level population is about 1.5 times lower, therefore,
the optical depth and the radius of the affected region should be 1.5--2 times
smaller (depending on the power-law index of the dust optical depth--frequency
dependence).
For the other sources this radius should be even smaller due
to lower optical depths and lower distances in some cases.


\begin{thebibliography}{}
\bibitem[2003]{aiwawa}
Aikawa, Y., Ohashi, N., Herbst, E., 2003, ApJ, 593, 906
\bibitem[1997]{bergin1}
Bergin, E. A., Goldsmith, P. F., Snell, R. L., Langer, W. D., 1997,
ApJ, 482, 285
\bibitem[1997]{bergin2}
Bergin, E. A., \& Langer, W. D., 1997, ApJ, 486, 316
\bibitem[1999]{bergin3}
Bergin, E. A., Plume, R., Williams, J. P., Myers, P. C., 1999, ApJ, 512, 724
\bibitem[1984]{botschwina}
Botschwina, P., 1984, Chem. Phys. Lett., 107, 535
\bibitem[2004]{bottinelli}
Bottinelli, S., \& Williams, J. P., 2004, A\&A, 421, 1113
\bibitem[1995]{braine}
Braine, J., Kr\"ugel, E., Sievers, A., Wielebinski, R., 1995, A\&A, 295, L55
\bibitem[1993]{brand}
Brand, J., \& Blitz, L., 1993, A\&A, 275, 67
\bibitem[1981]{cg}
Carroll, T. J., \& Goldsmith, P. F., 1981, ApJ, 245, 891
\bibitem[2005]{daniel}
Daniel, F., Dubernet, M.-L., Meuwly, M., Cernicharo, J., Pagani, L., 2005,
MNRAS, 363, 1083
\bibitem[1994]{doty}
Doty, S. S., \& Leung, C. M., 1994, ApJ, 424, 729
\bibitem[2004]{faundez}
Faundez, S., Bronfman, L., Garay, G., Chini, R., Nyman, L.-\AA., May, J.,
2004, A\&A, 426, 97
\bibitem[2004]{geppert}
Geppert W. D., Thomas, R., Semaniak, J., Ehlerding, A., Millar, T. J.,
\"Osterdahl, F., af Ugglas, M., Djuri\'c, N., Pa\'al, A., Larsson, M.,
2004, ApJ, 609, 459
\bibitem[2003]{}
Heninger, M., Lauvergnat, D., Lemaire, J., Boissel, P., Mauclaire, G.,
Marx, R., 2003, Int. J. Mass. Spectrom., 223-224, 669
\bibitem[1999]{jijina}
Jijina, J., Myers,  P. C., \& Adams,  F. C., 1999, ApJS, 125, 161
\bibitem[1996]{juvela96}
Juvela, M., 1996, A\&AS, 118, 191
\bibitem[1998]{juvela98}
Juvela, M., 1998, A\&A, 329, 659
\bibitem[1999]{lee1}
Lee, C. W., Myers, P. C., Tafalla M., 1999, ApJ, 526, 788
\bibitem[2001]{lee2}
Lee, C. W., Myers, P. C., Tafalla M., 2001, ApJS, 136, 703
\bibitem[2002]{li}
Li, Z.-Y., Shematovich, V. I., Wiebe, D. S., Shustov, B. M., 2002, ApJ, 569, 792
\bibitem[2005]{lin}
Lintott, C. J., Viti, S., Rawlings, J. M. C., Williams, D. A., Hartquist, T. W.,
Caselli, P., Zinchenko, I., Myers, P., 2005, ApJ, 620, 795
\bibitem[1997]{mardones}
Mardones, D., Myers, P. C., Tafalla M., Wilner D. J., Bachiller, R.,
Garay, G., 1997, ApJ, 489, 719
\bibitem[2000]{mccutcheon}
McCutcheon, W. H., Sandell, G., Matthews, H. E., Kuiper, T. B. H.,
Sutton, E. C., Danchi, W. C., Sato, T., 2000, MNRAS, 316, 152
\bibitem[1992]{meixner92}
Meixner, M., Haas, M. R., Tielens, A. G. G. M., Erickson, E. F.,
Werner, M, 1992, ApJ, 390, 499
\bibitem[1993]{meixner93}
Meixner, M., \& Tielens, A. G. G. M., 1993, ApJ, 405, 216
\bibitem[1998]{man}
Motte, F., Andr\'e, P., Neri, R., 1998, A\&A, 336, 150
\bibitem[1986]{mss}
Mozurkewich, D., Schwartz, P. R., Smith, H. A., 1986, ApJ, 311, 371
\bibitem[2002]{mueller}
Mueller, K. E., Shirley, Y. L., Evans II, N. J., \& Jacobson, H. R., 2002,
ApJS, 143, 469
\bibitem[1978]{neckel}
Neckel, T., 1978, A\&A, 69, 51.
\bibitem[2004]{nm}
Nomura, H., \& Millar, T. J., 2004, A\&A, 414, 409
\bibitem[2005]{ots}
Olmi, L., Testi, L., Sargent, A. I., 2005, A\&A, 431, 253
\bibitem[1994]{oh}
Ossenkopf, V., \& Henning T., 1994, A\&A, 291, 943
\bibitem[1986]{owrutsky}
Owrutsky, J. C., Gudeman, C. S., Martner, C. C., Tack, L. M.,
Rosenbaum, N. H., Saykally, R. J., 1986, J. Chem. Phys., 84, 605
\bibitem[1998]{pickett}
Pickett, H. M., Poynter, R. L., Cohen, E. A., Delitsky, M. L., Pearson, J. C.,
M\"uller, H. S. P., 1998, J. Quant. Spect. Rad. Trans., 60, 883
\bibitem[1998]{pz}
Pirogov, L. E., \& Zinchenko, I. I. 1998, AZh, 75, 14 (Astron. Rep., 42, 11)
\bibitem[2003]{pzcjm}
Pirogov, L., Zinchenko, I., Caselli P., Johansson,  L. E. B.,
Myers P. C., 2003, A\&A, 405, 639 (Paper~I)
\bibitem[1997]{plume}
Plume, R., Jaffe, D. T., Evans II, N. J., Martin-Pintado, J.,
Gomez-Gonzalez, J., ApJ, 476, 730
\bibitem[1976]{sk}
Scoville, N. Z., Kwan, J., 1976, ApJ, 206, 718
\bibitem[2003]{shem}
Shematovich, V. I., Wiebe D. S., Shustov, B. M., Li, Z.-Y., 2003, ApJ, 588, 894
\bibitem[2003]{shirley03}
Shirley, Y. L., Evans II, N. J., Young, K. E., Knez, C., Jaffe, D. T., 2003,
ApJS, 149, 375
\bibitem[2002]{tafalla02}
Tafalla, M., Myers, P. C., Caselli, P., Walmsley, C. M., Comito, C., 2002,
ApJ, 569, 815
\bibitem[1992]{turner}
Turner, B. E., Chan, K., Green S., Lubowich D. A., 1992, ApJ, 399, 114
\bibitem[1997]{ungerechts}
Ungerechts, H., Bergin, E. A., Goldsmith, P. F., Irvine, W. M., Schloerb, F. P.,
Snell, R. L., 1997, ApJ, 482, 245
\bibitem[2004]{dishoeck}
van Dishoeck, E. F., 2004, ARA\&A, 42, 119
\bibitem[1999]{wm}
Williams, J. P., \& Myers, P. C., 1999, ApJ, 518, L37
\bibitem[1994]{zin1}
Zinchenko, I., Forsstr\"om, V., Lapinov, A., \& Mattila, K., 1994, A\&A, 288, 601
\bibitem[1995]{zin3}
Zinchenko, I., Mattila, K., \& Toriseva, M., 1995, A\&AS, 111, 95
\bibitem[1998]{zin4}
Zinchenko, I., Pirogov, L., \& Toriseva, M., 1998. A\&AS, 133, 337
\end{thebibliography}
\end{document}